\documentclass[preprint,12pt,authoryear]{elsarticle}

\usepackage{amssymb}

\usepackage{color}
\usepackage{amsmath} 
\usepackage{bbold}
\usepackage{graphicx}
\usepackage{caption}
\usepackage{subcaption}

\newcommand{\Restspace}[1]{{\underline{\sigma}}_{#1}^{\rm spatial}}

\newcommand{\Resttime}[1]{{\underline{\sigma}}_{#1}^{\rm temporal}}
\newcommand{\lspace}[2]{\hat{#1}_{#2}}
\newcommand{\lspaceall}[2]{\overline{#1}_{#2}}
\newcommand{\sspace}[2]{\widetilde{#1}_{#2}}
\newcommand{\rsspace}[2]{\check{#1}_{#2}}

\newcommand{\dint}{{\rm \,d}}
\newcommand{\ltime}[1]{\left\langle #1\right\rangle}
\newcommand{\stime}[1]{\{ #1\}}

\newcommand{\ident}{\mathbb{1}}

\newcommand{\lslt}[2]{\ltime{\lspace{#1}{#2}}}
\newcommand{\lsltall}[2]{\ltime{\lspaceall{#1}{#2}}}
\newcommand{\lsst}[2]{\stime{\lspace{#1}{#2}}}

\newcommand{\rssst}[2]{\stime{\rsspace{#1}{#2}}}
\newcommand{\rsslt}[2]{\ltime{\rsspace{#1}{#2}}}
\renewcommand{\vec}[1]{\underline{#1}}
\newcommand{\mat}[1]{\underline{\underline{#1}}}

\journal{Journal of Computational Physics }

\definecolor{darkgreen}{rgb}{0.0, 0.5, 0.0}

\begin{document}

\begin{frontmatter}




\title{XLES Part II: From Extended Large Eddy Simulation to ODTLES}


\author[labelCottbus,labelFU]{Christoph Glawe}

\ead{Christoph.Glawe@protonmail.com}
\author[labelCottbus]{Heiko Schmidt}
\author[labelUSA]{Alan R. Kerstein}
\author[labelFU]{Rupert Klein}

\address[labelCottbus]{BTU Cottbus-Senftenberg, 
Siemens-Halske-Ring 14, 03046 Cottbus, Germany}

\address[labelUSA]{72 Lomitas Road, Danville, CA 94526, USA}

\address[labelFU]{FU Berlin, 
Arnimallee 6, 14195 Berlin,  Germany}

\begin{abstract}

%
%
%

In turbulence research and flow applications, turbulence models like RaNS 
(Reynolds averaged Navier-Stokes) 
models and LES (Large Eddy Simulation) are used. Both models filter the 
governing flow equations. Thus a scale separation approach is introduced for 
modeling purposes 
with the large scales simulated using a numerical scheme while smaller scales 
are assumed to be less important and might be modeled more or less easily.
Unfortunately 
small scales are frequently of big importance, e.g. in reactive flows, wall 
bounded flows, or flows with significant Prandtl or Schmidt 
number effects. 
Recent alternatives to these standard models are the class of models based on 
the one-dimensional turbulence (ODT) idea, like ODTLES. 
The ability of ODT to capture highly turbulent flows (recently up to 
$Re_{\tau}=6\times 10^{5}$) allows ODTLES to realize 3D resolutions 
basically independent of the turbulent intensity.
In two papers we provide a formal theory and application of an innovative 
modeling strategy 
for highly turbulent flows in domains of moderate complexity: 
%
In part I (see \cite{Glawe:2014}) a new general 
filtering approach, called XLES (extended LES), is introduced.
Contrary to LES, XLES is based on 2D filtering of the governing equations, 
whereby additional small scale terms are 
interpreted numerically.
In this work a new ansatz for the ODTLES model is introduced as one special 
approach in the XLES family of models by incorporating the ODT model into XLES. 
The ODT model introduces microstructures not captured by the XLES filtered 
equations. 
To illustrate the ODTLES model capabilities, turbulent channel and duct flows 
 up to friction Reynolds number 
$Re_{\tau} = 10000$ are studied.












  \end{abstract}

\begin{keyword}


Extended Large Eddy Simulations (XLES) \sep  
Large Eddy Simulation \sep One-Dimensional Turbulence \sep Stochastic 
Turbulence Modeling \sep Channel Flow 
\end{keyword}

\end{frontmatter}



\section{Introduction}
\label{s:Intro}

In fluid mechanics an increasing number of scientific and industrial problems 
are amenable 
to computer simulations due to growing computational power.
Realistic problems, e.g. in engineering and meteorology, require the 
application of turbulence models to reduce the computational effort. For an 
overview of turbulence properties and model approaches see e.g. 
\cite{Pope:2000}.

Large Eddy Simulation (LES, see e.g. \cite{Sagaut:2006}) is widely used in 
fundamental research and increasingly in industrial applications. In LES, 3D 
spatially filtered equations, containing large scale properties down to the 
inertial 
range of the turbulent cascade, are solved numerically. The effect of the 
unresolved scales is 
modeled, e.g. by an eddy viscosity model (see e.g. \cite{Germano:1991}).

In the area of reactive and wall bounded flows, especially with 
significant Prandtl or Schmidt number effects, small scale properties are very 
important, but not adequately represented by simple sub-grid models. 

Recent alternative models including the one-dimensional turbulence (ODT) model 
(see e.g. \cite{AR-Kerstein1999} and \cite{AR-Kerstein2001}) 
describe the 3D turbulence in a 1D sub-domain, whereby the 
numerical representation of molecular diffusive effects becomes computationally 
feasible also in highly turbulent flows. E.g. \cite{Meiselbach2015} presents 
wall bounded flows up to $Re_{\tau} = 6 \times 10^5$. This model appropriately 
describes individual flows involving one 
characteristic and predominant 
direction including the full turbulent cascade, which is valid inter 
alia in important problems within the research fields of fundamental combustion 
or atmospheric science.

In ODTLES the ODT model represents the microstructure terms (often called 
sub-grid model SGM), whereby the 
macrostructural model has to fulfill special requirements, which are 
satisfied by the extended Large Eddy 
Simulation (XLES) ansatz, introduced in part I (\cite{Glawe:2014}):
Three 2D filtered and coupled sets of equations are solved and discretized by 
three XLES-grids. Each 
set of equations corresponds to one predominant Cartesian direction.
The XLES unresolved sub-grid scale (SGS) terms are connected to 
the ODT advancement, where ODT is interpreted in terms of 
Navier-Stokes advection, rather than a stochastic (Monte-Carlo like) 
model.
This so called ODTLES model (2D filtered XLES equations with ODT 
microstructure modeling) combines the diffusive and 
advective small-scale effects computed by ODT with the ability to discretize 
3D domains within a macrostructural model. Because ODT is able to describe the 
full turbulent spectrum in a 1D sub-domain, only 3D 
effects not represented by ODT (e.g. the domain, secondary instabilities, ...) 
need to be resolved in 3D by XLES.

Here we distinguish between the expressions XLES to 
describe the XLES approach including an approximation or model for arising 
microscale terms, 
ODTLES if these terms are in particular described by ODT and XLES-U (XLES 
unclosed), if microscale terms are neglected. 

A previous version of ODTLES, introduced and examined by 
\cite{RC-Schmidt2010}, 
\cite{ED-Gonzalez-Juez2011}, and \cite{Glawe2013},
solves weakly coupled XLES equations that generate inconsistent 3D large 
scale velocity fields on the three XLES-grids. This introduces certain 
oscillations of the root mean square velocity (reported by 
\cite{ED-Gonzalez-Juez2011}) but no further major qualitative 
change in results. 


Though primarily intended to explain ODTLES to the 
LES community, XLES also gives new 
insights into ODTLES (e.g. allows to represent scalar properties 
by multiple XLES-grids consistently) and moreover is a novel autonomous 
modeling strategy because the ansatz is very general and not 
limited to one-dimensional models like ODT.

There are other approaches like LES-ODT (e.g. by \cite{Cao:2008}), LES-LEM 
(by \cite{Menon:2011}), and LEM3D 
(e.g. by \cite{Sannan:2013}) connecting 1D turbulence models and 3D 
computations. These approaches are not considered in detail in this work.

In this work the ODT model is briefly introduced in section \ref{s:ODT} and  
the XLES approach, introduced in detail in part I (\cite{Glawe:2014}), is 
summarized in section \ref{s:XLES}. Section \ref{s:XLES2ODTLES} introduces the 
ODT model as closure for XLES.
The time advancement cycle of the ODT closed XLES model (ODTLES) in 
section \ref{ss:TimeAdvancement} summarizes the model derivation, followed by 
channel and duct flow results in section \ref{s:Application}, comparative 
estimations of the computational costs in section \ref{s:XLES_ResProps}, and 
final conclusions in section \ref{s:Conclusions}.

In accord with part I (\cite{Glawe:2014}) we assume the flow to be described by 
the incompressible Navier-Stokes equations for a Newtonian fluid with constant 
kinematic viscosity ($\nu$) and constant density (for simplicity we assume 
$\rho_0 = 1$).

Note that in this work no Einstein summation convention is used.


\section{One-Dimensional Turbulence Model (ODT)}
\label{s:ODT}
\label{ss:ODTtimeAdvancement}

The ODT model describes the dynamics of a three dimensional turbulent flow 
within a one-dimensional sub-domain, including fully resolved molecular 
diffusion. Thus ODT is a dynamical model, able to describe e.g. the turbulent 
channel flow including high order flow 
statistics with high accuracy (see \ref{app:ODTResults}).

ODT stand-alone is able to compute meaningful results, even with one velocity 
component (see \cite{AR-Kerstein1999}). 
Nevertheless to capture anisotropic flow behavior and especially as a closure 
within a 3D approach, two or three velocity components are advantageous.  

In wall-bounded flows (especially in turbulent channel flows) described by 
ODT with 
three velocity components, the wall-normal and the 
spanwise velocities are identical, so two-component ODT captures similar 
statistical flow properties. (Note that ODT results in 
\ref{app:ODTResults} are computed with 2 velocity components). 

In this section the ODT time advancement is 
described briefly.
Here we introduce ODT including $2$ velocity components. This ODT model is a 
modification of the ODT vector formulation by \cite{AR-Kerstein2001} including 
$3$ velocity components.


ODT emulates the time evolution of a turbulent 3D fluid in a 1D subspace, which 
is oriented in the Cartesian $x_k$-direction. The $2$ ODT velocity components 
$u_{k,i}$ (with $k\neq i$) are oriented orthogonally to the $x_k$-direction.  
The time evolution of a velocity field $u_{k,i}$ in this 1D 
subspace is described by:
	\begin{align}
	\label{eqn:ODTtimeEvolution}
		\left( 
		  \partial_t { u_{k,i}} + \mathcal{D}_{\rm 
		  ODT_k}({ 
		  u_{k,i}}) + e_{k,i}(u_{k,i};x_0,l)  
		\right)  
		  = 0 \,\, {\rm with} \,\, k,i=\{1,2,3\}\wedge i \neq k
	\end{align}
with the ODT diffusion term $\mathcal{D}_{\rm ODT_k}({ 
u_{k,i}}) = - \nu \partial_{x_k}^2 { u_{k,i}}$ which is numerically approximated
by an implicit Euler scheme in time and a central difference scheme in space. 
The 
index notation resembles the XLES index notation used in the following sections 
with the velocity $u_{k,i}$ oriented in the $x_i$-direction within a 1D 
sub-space oriented in the $x_k$-direction ($i \neq k$).


The term  $e_{k,i}(u_{k,i};x_0,l)$ is an 
instantaneous eddy function affecting ${u_{k,i}}$ within the eddy 
range $x_k\in[x_0,x_0+l]$. 
The maximum eddy length $l^{\rm max}$ is enforced, hence $l \leq l^{\rm max}$. 
The eddy function $e_{k,i}$ is introduced to represent 
a stochastic procedure that emulates turbulent advection:
	\begin{align}
	e_{k,i} : 	{ u_{k,i}}(x_k,t) \to  
{u_{k,i}}(f(x_k,l),t) +  {c}_i K(x_k)  
	\label{eqn:eddyfunction}.
	\end{align}
Note that the advection function $e_{k,i}$ depends on both 
velocity components $u_{k,i}$ (with $i=\{1,2,3\} \wedge i \neq k$) due to $c_i$ 
in Eq. (\ref{eqn:ODT_KernelAmplitude}).
The mapping function $f(x_k,l)$, representing fluid transport, is 
measure preserving (the non-local analog of vanishing velocity divergence), 
continuous, and satisfies the requirement of scale locality (at most 
order-unity 
changes in property gradients). These  indispensable physical requirements for 
$f(x_k,l)$ are satisfied by a triplet map, which places three compressed copies 
of the original profile $\{{
u_{k,i}}(x_k) , \,x_k \in [x_0, x_0+l] \}$  in the eddy 
range. The middle copy is reversed to preserve continuity. 
The triplet map $f(x_k,l) \to x_k$ is:
	\begin{align}
		f(x_k,l) &= x_0 + 
		\begin{cases}
			3(x_k-x_0), 		& \mathrm{if} \;\;x_0 \le x_k 
\le x_0 + \frac{1}{3} l \\
			2l - 3(x_k-x_0), 	& \mathrm{if} \;\;x_0 
+\frac{1}{3} l  \le x_k \le x_0 + \frac{2}{3} l \\
			3(x_k-x_0) -2 l, 	& \mathrm{if} \;\;x_0 
+\frac{2}{3} l  \le x_k \le x_0 +  l \\
			(x_k-x_0), 		& \mathrm{else}.
		\end{cases}
		\label{eqn:mapping} 
	\end{align}

In Eq. (\ref{eqn:eddyfunction}), $K(x_k)$ is a kernel function which in 
combination with the amplitudes $c_i$ assures momentum and energy 
conservation and controls the energy redistribution among the velocity 
components.
A possible definition is: $K(x_k) = x_k - f(x_k,l)$.
This energy redistribution is a 1D interpretation of the pressure-fluctuation 
effect in a 3D flow and therefore is called pressure scrambling.

Determination of the amplitudes $c_i$ requires additional modeling.

\cite{AR-Kerstein2001} derive the amplitudes:
\begin{align}
\label{eqn:ODT_KernelAmplitude}
  c_i = \frac{27}{4 l} 
  \left(
     - u_{K;k,i} + \textrm{sign}(u_{K;k,i})  
     \sqrt{u_{K;k,i}^{2} + \sum_{j} \alpha T_{ij} u_{K;k,j}^{2} }
  \right)\,;\, i\neq k,\, j\neq k
\end{align}
with the definition
\begin{align}
 u_{K;k,i} \equiv \frac{1}{l^2} \int u_{k,i} (f(x_{k})) K(x_k) \dint x_k
\end{align}
and the transfer matrix
\begin{align}
 \label{eqn:2VelCompTransMat}
  \alpha T = \alpha
  \begin{pmatrix}
      -1 & 1 \\
       1 &-1 \\
  \end{pmatrix}.
\end{align}
%
%
The free parameter $\alpha$ ensures the amplitudes $c_i$ in Eq. 
(\ref{eqn:ODT_KernelAmplitude}) to be real values for $0\leq \alpha \leq 1$. We 
choose $\alpha=1/2$ corresponding to the equalization of the two component 
available energies in the present formulation. 

During an eddy event, the transfer matrix $T$ redistributes the turbulent 
kinetic energy among velocity components. 

This `pressure scrambling' accounts for the tendency for pressure 
fluctuations to restore isotropy and is 
invariant under exchange of indices. 
By construction the momentum and total energy are not changed by the pressure 
scrambling.

%
%

During the ODT time evolution in Eq. (\ref{eqn:ODTtimeEvolution}), the eddy 
size $l$ and the location $x_0$ are sampled from a probability distribution 
representing the physics. For a given $\{l,x_0\}$ an eddy turnover time 
$\tau_{e}$ can be calculated leading to an occurrence frequency 
$\frac{1}{\tau_e}$. Since the ODT 
triplet map is an instantaneous process, the frequency for the eddy specified 
by 
$\{l,x_0\}$ is chosen from an event rate distribution:
	\begin{equation}
		\lambda(x_0,l) = \frac{C}{l^2 \tau_e(x_0,l)} = 
\frac{C}{l^3} 
\sqrt{E_{kin} - E_{pot} - \frac{\nu^2}{l^2} Z  }
		\label{eqn:eventrate}		
	\end{equation}
involving particular definitions of the turbulent kinetic energy $E_{kin}$ and 
the  
potential energy $E_{pot}$. The latter vanishes for the cases considered in 
this work. 
Note that the work by \cite{WT-Ashurst2005} introduces a variable density 
formulation of ODT including $E_{pot}$. 
The values $l^{\rm max}$, $C$, and $Z$ are adjustable model parameters. 
The latter is introduced to cut off eddies with unphysically small energy and 
the parameter $C$ is an overall rate coefficient determining the strength of 
the turbulence. 
The maximum eddy length $l^{\rm max}$ is chosen to characterize the 
largest (global) scale within the flow, e.g. the channel half 
height.

\section{Extended Large Eddy Simulation (XLES)}
\label{s:XLES}

In sections \ref{ss:XLES_scales}, \ref{ss:XLES_momentum}, 
and \ref{ss:XLES_mass} the XLES framework is summarized briefly, to allow 
focus on features related to ODTLES without repeating details introduced 
in part I (\cite{Glawe:2014}).

Additionally a time scale separation is introduced in section 
\ref{s:TimeSeparation} to simplify the interpretation of microscale terms with 
reference to the ODT advancement.

\subsection{XLES: Spatial Filtering}
\label{ss:XLES_scales}

\begin{figure}
        \centering
        \begin{subfigure}[b]{0.26\textwidth}
                \includegraphics[width=\textwidth]{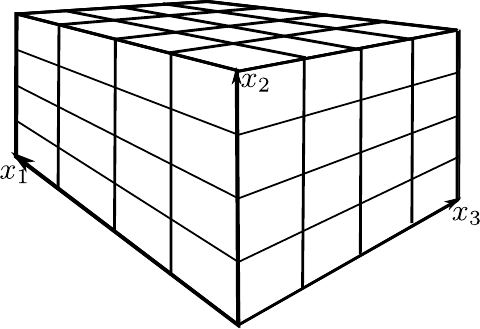}
                \caption{$[l_1 l_2 l_3]u_i=\lspaceall{u}{i}^{\rm LES}$.}
                \label{fig:gridLES}
        \end{subfigure}        
        \begin{subfigure}[b]{0.23\textwidth}
                \includegraphics[width=\textwidth]{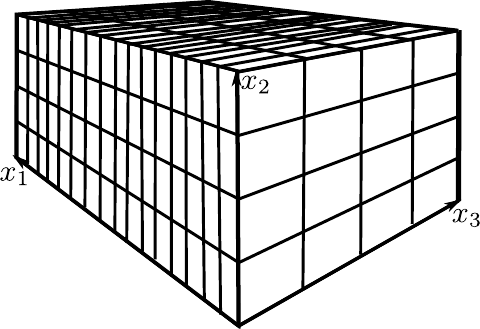}
                \caption{grid 1: for $\lspace{u}{1,i}$.}
                \label{fig:grid1}
        \end{subfigure}
        \begin{subfigure}[b]{0.23\textwidth}
                \includegraphics[width=\textwidth]{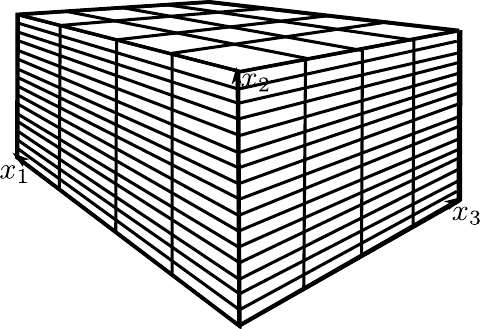}
                \caption{grid 2: for $\lspace{u}{2,i}$.}
                \label{fig:grid2}
        \end{subfigure}
	\begin{subfigure}[b]{0.23\textwidth}
                \includegraphics[width=\textwidth]{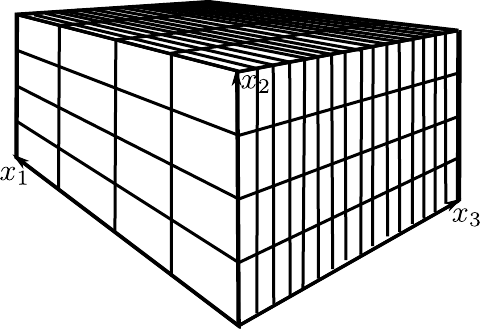}
                \caption{grid 3: for $\lspace{u}{3,i}$.}
                \label{fig:grid3}
        \end{subfigure}
        \caption{In XLES the velocity components ${u}_i$ are resolved using 
multiple 
XLES-grids illustrated in \ref{fig:grid1}-\ref{fig:grid3}. 
3D large scale properties, corresponding to a standard LES grid with 
(illustrative) $N_{\rm 
LES}=4$ cells, are illustrated in 
\ref{fig:gridLES}.
Illustrative XLES-resolved small scale (`RSS') properties are approximated 
using 
$N_{\rm RSS}=16$ 
 cells in \ref{fig:grid1}-\ref{fig:grid3}. }\label{fig:grids}
\end{figure}

The basic XLES concept is to apply 2D filters to the 
velocity field, maintaining
one Cartesian direction highly resolved.
Using these filters, three 2D filtered velocity fields, each 
corresponding to one highly resolved Cartesian direction, are introduced and 
discretely 
represented by three staggered XLES-grids, illustrated in figure 
\ref{fig:grid1} -- \ref{fig:grid3}.

A vector-matrix notation (which we will refer to as XLES-vector notation) is 
used 
to represent the three 2D filtered velocity fields:
\begin{align}
\label{eqn:2dfilter}
  \begin{pmatrix}
   [l_2 l_3]{u}_{i} \\
   [l_1 l_3]{u}_{i} \\
   [l_1 l_2]{u}_{i}
 \end{pmatrix}
  = 
  \begin{pmatrix}
	l_2 l_3 & 0 & 0 \\ 
	0 & l_1 l_3 & 0 \\ 
	0 & 0 & l_1 l_2 
    \end{pmatrix}
    \begin{pmatrix}
	{u}_{i} \\
	{u}_{i} \\
	{u}_{i}
   \end{pmatrix}
  \equiv
     \mat{l}^{2D} \, \vec{u}_i
  \equiv   
     \lspace{\vec{u}}{i} , \;
     {\rm with} \, i=\{1,2,3\} 
\end{align}
where $i$ indexes the velocity components. We introduce another index $k$ 
referring to the velocity field $\lspace{u}{k,i}$ represented in XLES-grid $k$ 
(similar to the index notation in section \ref{s:ODT}).

Here  without loss of generality (w.l.o.g.) the 2D filter operator $[l_2 l_3]$ 
(corresponding to XLES-grid 
$1$) represents a tensor product of 1D filter 
operators $[l_2]$ and $[l_3]$. This tensor product is commutable: $[l_2 
l_3]$=$[l_3 l_2]$.

The 1D filter operators are used to derive micro-and macroscale velocity
terms:
\begin{align}
\label{eqn:XLESscaleAnsatz}
       u_i
    =& \left[  {  l_1 l_2 l_3 +s_1 l_2 l_3 + l_1 s_2 l_3 + 
l_1 l_2 s_3} \right] u_i \\
    +& \underbrace{\left[ 
 s_1 s_2 l_3 + s_1 l_2 s_3 + l_1 s_2 s_3 + s_1 s_2 s_3 \right]  
}_{\equiv \mathcal{S}\, (\rm denoting \;XLES \;SGS \;terms)} u_i.
\nonumber 
\end{align}
with the 1D small scale operator $s_k = \ident -l_k$ ($k=\{1,2,3\}$) and the 
unity operator $\ident$.

These scales are decomposed (using the 2D filter 
matrix $\mat{l}^{2D}$): 
\begin{align}
  \label{eqn_XLES_AllScalesDef}
   \vec{u}_i 
   = 
     \begin{pmatrix}
   [l_2 l_3] {u}_{i} \\
   [l_1 l_3] {u}_{i} \\
   [l_1 l_2] {u}_{i}
 \end{pmatrix} + 
 \begin{pmatrix}
   [ l_1 s_2 l_3 +l_1 l_2 s_3 ]u_i \\
   [ s_1 l_2 l_3 +l_1 l_2 s_3 ]u_i \\
   [ s_1 l_2 l_3 +l_1 s_2 l_3 ]u_i
 \end{pmatrix}
 + \begin{pmatrix}
   [ \mathcal{S}]u_i \\
   [ \mathcal{S}]u_i \\
   [ \mathcal{S}]u_i
 \end{pmatrix}
  \equiv  \mat{l}^{2D}\, \vec{u}_i +  \mat{C} \,\mat{s}^{1D} \mat{l}^{2D} 
\vec{u}_i + [ 
\mathcal{S}]\vec{u}_i
\end{align}
into three terms:
\begin{enumerate}
\item `Directly Resolved': 
 
  $\mat{l}^{2D}\, \vec{u}_i$ is discretely represented in the XLES-grids.
\item `Indirectly Resolved':

  The resolved small scale (RSS) velocities 
  \begin{align}
   \label{eqn:Mom_ResolvedSmallScales}
  \rsspace{\vec{u}}{i} \equiv  \mat{s}^{1D}  \mat{l}^{2D} \vec{u}_i
  = \lspace{\vec{u}}{i} - \lspaceall{\vec{u}}{i}^{\rm LES}
  \end{align}
  (with $\lspaceall{\vec{u}}{i}^{\rm LES} = [l_1 l_2 l_3] \vec{u}_i$)
  lead to coupling terms between the XLES-grids, and are directly resolved 
by another XLES-grid (see section \ref{ss:XLES_momentum} for details).
The coupling matrix $\mat{C}$ and the small scale matrix $\mat{s}^{1D}$ are:
  \begin{align}
  \mat{C}    = 
      \begin{pmatrix}
	0 	& \ident	& \ident \\
	\ident & 0	 	& \ident \\
	\ident & \ident 	& 0
    \end{pmatrix}
  \, {\rm and } \quad 
    \mat{s}^{1D} =
      \begin{pmatrix}
	s_1 		& 0		 &	0 \\
    0			& s_2	 	 &	0 \\
      0			& 0		 & 	s_3
    \end{pmatrix}.
  \end{align}
The LES velocity field   $\lspaceall{{u}}{i}^{LES}$ corresponds to the 1D 
filtered XLES velocity field:  $[l_k]\lspace{u}{k,i}$ (see figure 
\ref{fig:grids}).

\item `Not Resolved':

\begin{align}
 \label{eqn:XLES_unresolvedVel}
 [\mathcal{S}]u_i = [s_1 s_2 l_3 + s_1 l_2 s_3 + l_1 s_2 
s_3   + s_1 s_2 s_3]u_i \equiv \sspace{u}{i}
\end{align}
is not resolved in any 
XLES-grid and leads to 
the XLES microscale terms. 
\end{enumerate}

Because the microscale model (ODT) is able to represent large scale effects and 
the 
macroscale model (XLES) contains 1D small scale terms, the classical 
term `scale separation' is misleading for XLES and especially ODTLES. Thus 
we will refer to a `filter separation'.


\subsection{XLES: Momentum Conservation}
\label{ss:XLES_momentum}

The 2D filtered XLES momentum equations are 
\begin{align}
 \label{eqn:XLES_PreFilMomentumGridN2}
  0 =& \partial_{x_i}  
\vec{\lspaceall{p}{}}^{\rm LES} +  \left( \partial_t  - \nu 
\sum_{j=1}^3 
\partial_{x_j}^2   \right) 
\vec{\lspace{u}{}}_{i}
+  \sum_{j=1}^{3} \partial_{x_j}  
\vec{\lspace{u}{}}_{j}* \vec{\lspace{u}{}}_{i}\\
  +&  \sum_{j=1}^{3} 
   \left(
	 \mathcal{\vec{R}}_{ij}^{\rm XLES} + \mathcal{\vec{C}}_{ij}^{\rm XLES}
	+\mathcal{\vec{X}}_{ij}^{XLES}
   \right)
   + \Restspace{} \nonumber
\end{align}
with the entry-wise multiplication $*$ between XLES-grid vectors (and matrices).
The decomposed SGS Reynolds stresses $\mathcal{\vec{R}}_{ij}^{\rm XLES}$ and 
the cross-stress terms $\mathcal{\vec{C}}_{ij}^{\rm XLES}$ are combined to 
obtain the XLES sub-grid scale (SGS) advection terms:
\begin{align}
\label{eqn:XLES:ModelTerm}
  \mat{l}^{2D}\mathcal{\vec{M}}_{ij} \equiv 
  \mat{l}^{2D}\partial_{x_j}  
 \left(
	\sspace{\vec{u}}{j} * \rsspace{\vec{u}}{i} 
      + \rsspace{\vec{u}}{j}* \sspace{\vec{u}}{i}
      + \frac{1}{2} 
      \left(
	      \sspace{\vec{u}}{j} 	\lspaceall{\vec{u}}{i}^{\rm LES}
	  +   \lspaceall{\vec{u}}{j}^{\rm LES}\sspace{\vec{u}}{i}
	  +   \sspace{\vec{u}}{i}	\sspace{\vec{u}}{j}
      \right)
  \right) 
\end{align}
and its coupling $ \mat{l}^{\dag}* \mat{C}  \, \mat{l}^{2D} 
\mathcal{\vec{M}}_{ij}$. Here $\sspace{\vec{u}}{i}$ is the vector notation for 
Eq. (\ref{eqn:XLES_unresolvedVel}), $\rsspace{\vec{u}}{i}$ is defined in 
Eq. (\ref{eqn:Mom_ResolvedSmallScales}), and $\mat{l}^{\dag}$ is defined in Eq. 
(\ref{eqn:FilterDag}).

The linearized coupling stress terms of the resolved small scale velocities 
$\lspace{\vec{u}}{i}$ are:
\begin{align}
 \mathcal{\vec{X}}_{ij}^{XLES} = 
\mat{l}^{\dag}* \partial_{x_j}  \mat{C}
 \left(
	  \lspace{\vec{u}}{j} 		* \lspace{\vec{u}}{i}
	- \lspaceall{\vec{u}}{j}^{\rm LES} 	* \lspaceall{\vec{u}}{i}^{\rm 
LES}
\right) ,
\end{align}
which has a discrete representation. 
Hereby the matrix $\mat{l}^{\dag}$ with
\begin{align}
\label{eqn:FilterDag}
    \mat{l}^{\dag} = 
    \begin{pmatrix}
    \ident & {l_1^{-1}}{l_2} & {l_1^{-1}}{l_3} \\
    {l_2^{-1}}{l_1} & \ident & {l_2^{-1}}{l_3} \\
    {l_3^{-1}}{l_1} & {l_3^{-1}}{l_2} & \ident 
    \end{pmatrix} 
\end{align}
includes the deconvolution operator $l_k^{-1}$ which reconstructs filtered 
information. Due to the coupling the large scale influenced by highly resolved 
effects in a specific XLES-grid is communicated to the other XLES-grids.
Hereby the reconstruction is executable because the operator $[l_k^{-1}]$ is 
only applied to large scale terms in $x_k$-direction.
For details about the coupling procedure and an possible
algorithm for a discrete deconvolution, please see \cite[section 
3.1]{Glawe:2014}.


The spatial XLES model error terms are:
\begin{align}
\label{eqn:XLES:SpatialModelError}
\Restspace{} =& 
 \partial_{x_j} \mat{l}^{2D} \vec{1} 
   (
      \rsspace{u}{1,j} \rsspace{u}{2,i}
    + \rsspace{u}{1,j} \rsspace{u}{3,i} 
    + \rsspace{u}{2,j} \rsspace{u}{1,i} 
    + \rsspace{u}{2,j} \rsspace{u}{3,i} 
    + \rsspace{u}{3,j} \rsspace{u}{1,i} 
    + \rsspace{u}{3,j} \rsspace{u}{2,i}
   )
   +
    \mathcal{\vec{L}}_{ij}^{2D}
\end{align}
including non-linear coupling terms and the 2D Leonard 
stresses $\mathcal{\vec{L}}_{ij}^{2D}$.

\subsection{XLES: Time Scale Separation}
\label{s:TimeSeparation}

The XLES advection terms are represented by three overlapping 
XLES-grids, including coupling terms between these XLES-grids, and 
an additional ODT advancement for ODTLES. 
This ODT advancement involves 
instantaneous stochastic mappings whose instantaneous nature is in conflict 
with the idea of high order time integration schemes. 

On the one hand, the simplest and physically most convenient way to 
advance the coupling terms and the dynamical SGM within an XLES framework is an 
explicit Euler scheme, which allows a straightforward interpretation of 
coupling terms and the stochastic turbulent advection within ODT. 

On the other hand, an efficient numerical advection 
scheme includes high order time integration, which is even required for 
stability reasons by some spatial discretizations. 

A known compromise is to linearize the advection:
The linear advection part is advanced by a high order numerical 
scheme (details in part I (\cite{Glawe:2014})), while the non-linear part is 
implemented by a 1st order explicit Euler scheme.
One possible way to interpret such an approach is to integrate the dynamical 
velocity field over one time step and use this velocity field to advect the 
dynamical variables within the next time step. This approach can easily include 
random occurrences of instantaneous ODT mappings and therefore is suitable for 
ODTLES. 

In part I (\cite[section 3.3]{Glawe:2014}), alternative time schemes are 
suggested which potentially avoid the linearization of the advection 
terms. 

In contrast to RaNS models, time averaging is not applied to the dynamical 
variables, but to the advecting velocity variables. This is especially 
reasonable because ODT is a dynamical model introducing small time scale 
effects. 

Formally a time scale separation is invoked that simplifies the 
inclusion of ODT 
into XLES, though this procedure is not required by XLES (in part I 
(\cite{Glawe:2014})) time scales are not separated).
The separated time scales are:
\begin{align}
 \label{eqn:TimeSeparation}
  u_j = \ltime{u_j} + \stime{u_j}
\end{align}
with the large time scale
\begin{align}
  \ltime{u_j} = \frac{1}{\tau} \int_{t}^{t+\tau}  {u_j} \dint t'
\end{align}
and the small time scale (fluctuations) $\stime{u_j} = u_j - \ltime{u_j}$. 
Note that the time filter and spatial filters in XLES are 
independent of each other:$\ltime{[l_k]u_{j}} = [l_k] \ltime{u_j}$. 

The integral time $\tau$ corresponds to the 3D large scale flow. 
A natural choice for $\tau$ is the discrete 
time step size of the 3D large scale advancement scheme.
Thus the modeling strategy is directly connected to the numerical realization.

If an explicit time integration scheme is used, $\tau$ is numerically 
restricted by a 
(global) 
Courant-Friedrichs-Lewy condition. Here $CFL$ defines the constant $CFL$-number 
depending on the implemented numerical scheme:
\begin{align}
\label{eqn:XLES_CFL}
    \tau = CFL \, \min_{k,i} \left( \frac{\Delta x_k^{\rm 
LES}}{\lspace{u}{k,i}} 
\right)
\end{align}
within all XLES-grids $k=\{1,2,3\}$ and with all velocity directions $x_i$ ($
i=\{1,2,3\}$. Local time step restrictions are possible but not implemented.

The time scale separation within XLES-U implies additional error terms
caused by time scale separation:
 \begin{align}
   \label{eqn:XLES:TemporalErrorXLESU}
    \Resttime{XLES-U} &=  \sum_{j=1}^3  
      \partial_{x_j}\lsst{\vec{u}}{1,j} \lspace{\vec{u}}{1,i}.
 \end{align}
 
By defining the integral time scale $\tau$ based on the resolved small scale 
cell size $\Delta x_k^{\rm RSS}$ in XLES-grid $k$, the time scale separation 
is suppressed, because all XLES-U velocities are large scale in time:
\begin{align}
\label{eqn:XLES_SmallTau}
      \tau = CFL \, \min_{k,i} \left( \frac{\Delta x_k^{\rm 
RSS}}{\lspace{u}{k,i}} \right)
\end{align}
The $CFL$-number can be used to switch between Eq. (\ref{eqn:XLES_CFL}) and Eq. 
(\ref{eqn:XLES_SmallTau}) and thus becomes a model parameter balancing (and 
controlling) the 
temporal model error $\Resttime{XLES-U}$ and the model performance.

Thus the $CFL$-number can be increased for performance reasons: ODTLES 
is still stable and well defined for $CFL=0.5 \frac{N_{\rm RSS}}{N_{\rm LES}}$ 
which e.g. for $N_{\rm RSS}= 512$ and $N_{\rm LES}=16$ leads to a factor $32$ 
increased XLES time-step size (ODT advancement is indirectly influenced) with 
additional model errors (see Eq. (\ref{eqn:XLES:TemporalErrorXLESU})) having 
only a small impact on specific problems. In this work all 
results are computed with 
considerably small $CFL$-number, following Eq. (\ref{eqn:XLES_SmallTau}).

Since ODTLES allows a huge number of turbulent ODT events within 
the time scale $\tau$, the averaged velocities are smoothed, which 
is part of the ODTLES modeling strategy. Again the $CFL$-number is controlling 
this modeling impact, because with $\tau$ based on Eq. 
(\ref{eqn:XLES_SmallTau}) the number of ODT turbulent events is decreased.

ODT describes fluctuations (small time scale terms) in 
ODT-direction $x_{k}$ corresponding to advection terms of the form $\partial_j 
\lsst{\vec{u}}{k,k} 
\lspace{{u}}{k,i}$. 
The corresponding time averaged advecting 
velocity $\lslt{u}{k,k}$ is specified due to mass conservation, see section 
\ref{ss:XLES_mass}.  
The terms $\partial_j 
\lsst{\vec{u}}{k,k} 
\lspace{{u}}{k,i}$ correspond formally to the spatial 
XLES macroscale (they are part of the model error in Eq. 
(\ref{eqn:XLES:TemporalErrorXLESU}), controlled by $CFL$), but can be 
interpreted by ODT, which is possible due to the time scale 
separation.

Additionally to the terms $\mathcal{\vec{M}}_{ij}$ ODT directly and indirectly 
(due to coupling) represents fluctuations across 3D large scale cells:
%
 \begin{align}
  \label{eqn:ODT_STimeAdvTerms}
 \sum_{j=1}^3 
  \left(
  \partial_{x_j}  \lsst{\vec{u}}{j}* \lspace{\vec{u}}{i}
  +(
   \mat{C}   
\partial_{x_j}  \lsst{\vec{u}}{j}* \lspace{\vec{u}}{i} )^T
  \right)
=  \left( \mat{\ident} + \mat{l}^{\dag} * \mat{C}\, \right) 
  \begin{pmatrix}
     \partial_{x_1}\lsst{u}{1,1} \lspace{u}{1,i} \\
     \partial_{x_2}\lsst{u}{2,2} \lspace{u}{2,i} \\
     \partial_{x_3}\lsst{u}{3,3} \lspace{u}{3,i}
  \end{pmatrix}
  + \Resttime{ODTLES}.
 \end{align}
This model assumption leads to a model error term replacing Eq. 
(\ref{eqn:XLES:TemporalErrorXLESU}):
 \begin{align}
   \label{eqn:XLES:TemporalError}
    \Resttime{ODTLES} &= 
      \begin{pmatrix}
	  \partial_{x_2} \rssst{u}{1,2} \rsspace{u}{1,i} 
	+ \partial_{x_3} \rssst{u}{1,3} \rsspace{u}{1,i}  \\
	  \partial_{x_1} \rssst{u}{2,1} \rsspace{u}{2,i} 
	+ \partial_{x_3} \rssst{u}{2,3} \rsspace{u}{2,i}  \\
	  \partial_{x_1} \rssst{u}{3,1} \rsspace{u}{3,i} 
	+ \partial_{x_2} \rssst{u}{3,2} \rsspace{u}{3,i}  
      \end{pmatrix}.
 \end{align}
The error term in Eq. (\ref{eqn:XLES:TemporalError}) summarizes 
all fluctuating terms not in ODT direction $x_k$. Again this error term is 
controlled by the $CFL$-number.

Applying the time scale separation to the advecting velocities leads to a 
modified 
XLES momentum equation (compare to Eq. (\ref{eqn:XLES_PreFilMomentumGridN2})):
\begin{align}
 0 =& \partial_{x_i}  
{\lslt{\vec{p}}{}}^{} +  \left( \partial_t 
- \nu 
\sum_{j=1}^3 
\partial_{x_j}^2   \right) 
\vec{\lspace{u}{}}_{i}
+  \sum_{j=1}^{3} \partial_{x_j}  
{\lslt{\vec{u}}{j}}* 
\vec{\lspace{u}{}}_{i} \nonumber \\
 \label{eqn:XLES_PreFilMomentumGridN}
  +&\mathcal{\vec{M}}_{ODT} + \Restspace{}  + \Resttime{} \\
   +&  \sum_{j=1}^{3} 
      \mat{l}^{\dag}* \partial_{x_j}   \mat{C}
      \left(
		\lslt{\vec{u}}{j} 		* \lspace{\vec{u}}{i}
	      - \lsltall{\vec{u}}{j}^{\rm LES} 	* 
\lspaceall{\vec{u}}{i}^{\rm LES} \right) 
   +  \mat{l}^{\dag}* 
   \mat{C}\,  \mathcal{\vec{M}}_{ODT}  \nonumber .
\end{align}
The advection terms assumed to be modeled directly by ODT (involving $3$ 
velocity components) are:
\begin{align}
 \label{eqn:ODTadvectionIdeal}
  \mathcal{\vec{M}}_{ODT} =   \sum_{j=1}^{3} 
   \left( \mat{l}^{2D}
	 \mathcal{\vec{M}}_{ij}^{\rm XLES} + 
\partial_{x_i} \lsst{\vec{p}}{}
	+ \partial_{x_j}  {\lsst{\vec{u}}{j}}* 
\vec{\lspace{u}{}}_{i} 
   \right) .
\end{align}
The derivation of $\mathcal{\vec{M}}_{ODT}$ is 
tailored for ODT, which is emphasized by the acronym ODT.
Additionally the ODT advancement is
coupled between the XLES-grids ($\mat{l}^{\dag}* 
   \mat{C}\,  \mathcal{\vec{M}}_{ODT}$).

Note that spatial large scale terms of the form 
$\partial_{x_2} \stime{\lspaceall{u}{1,2}^{\rm LES}} \lspaceall{u}{1,i}^{\rm 
LES}$ are included in the ODT coupling $\mat{l}^{\dag}* \mat{C}\,  
\mathcal{\vec{M}}_{ODT}$.

The term $\partial_{x_i}  \lsst{\vec{p}}{}= \mat{l}^{2D}
\partial_{x_i}{\stime{\vec{p}}}$ describes pressure fluctuations.
By construction ODT is mass conservative, nevertheless pressure fluctuations
are modeled by applying the so called pressure scrambling 
(see section \ref{ss:ODTtimeAdvancement}).

The full ODTLES advancement cycle is summarized in section 
\ref{ss:TimeAdvancement}.


An alternative approach, applied by \cite{Cline:2015} within the 
lattice-based multiscale simulation model (LBMS), is to 
couple each individual turbulent event within the ODT advancement, instead of 
the time averaging approach introduced here. 
This approach potentially introduces small time scale communication within a 
parallel algorithm.

\subsection{XLES: Mass Conservation}
\label{ss:XLES_mass}

In the incompressible flow regime, the 2D filtered velocity fields need to be 
divergence free to ensure mass conservation. 

Because the XLES dynamics take place on the integral time scale ($\tau$), the 
2D 
filtered mass equation 
\begin{align}
\label{eqn:XLES_massTime}
 0 &= \sum_{i=1}^3 \partial_{x_i} \mat{l}^{2D}  \ltime{\vec{u}}_{i}
=
  \sum_{i=1}^3 \partial_{x_i} \lsltall{\vec{u}}{i}^{\rm LES}  +   
  \sum_{i=1}^3 \partial_{x_i} \rsslt{\vec{u}}{i}	   
\end{align}
is enforced for the integral time scale by the procedure described 
in this section, while 
velocity fluctuations (corresponding to the small time scale), described by the 
ODT advancement, are mass conservative by construction (see section 
\ref{s:ODT}). 

The mass conservation of the 3D large scale velocity $\lsltall{u}{i}^{\rm 
LES}$ is enforced by a standard 
approach: A pressure Poisson equation is solved, leading to a large scale 
pressure field $\lsltall{\vec{p}}{}^{\rm LES}$ (see standard textbooks, e.g. 
\cite{Ferziger1999}). The resulting pressure gradient $\partial_{x_i} 
\lsltall{\vec{p}}{}^{\rm LES}$ enforces a divergence free velocity field 
$\lsltall{u}{i}^{\rm LES}$ by solving: $\partial_t  \lsltall{u}{i}^{\rm LES} + 
\partial_{x_i} 
\lsltall{\vec{p}}{}^{\rm LES} = 0$  (see section \ref{ss:TimeAdvancement} and 
especially Eq. (\ref{eqn:timeIntegral})).
Here a consistent 3D large scale field $\lsltall{u}{i}^{\rm LES} = 
[l_1] \lslt{u}{1,i} = [l_2] \lslt{u}{2,i} = [l_3] \lslt{u}{3,i}$ is required, 
which is enforced by the coupling terms $\ltime{\mathcal{\vec{X}}_{ij}^{\rm 
XLES}}$ and 
$\ltime{\mat{l}^{\dag} * \mat{C}\, \mathcal{\vec{M}}_{\rm ODT} }$.

The resolved small scale (RSS) divergence $\sum_{i=1}^3 \partial_{x_i}  
\rsslt{\vec{u}}{i}=0$ 
vanishes if a box filter $[l_k]\ltime{u_i} = \frac{1}{\Delta 
x_k} \int_{-\frac{\Delta 
x_k}{2}}^{\frac{\Delta 
x_k}{2}}  \ltime{u_i} \dint x_k'$ is used (proof in 
(\cite[Appendix B]{Glawe:2014})).

Since the 3D velocities $\lsltall{\vec{u}}{j}^{\rm LES}$ are divergence free 
(after solving the pressure Poisson problem), a 
direct solver can compute one velocity component $\lslt{u}{k,k}$ in each 
XLES-grid $k$ within one 3D cell of the size $\Delta x_k$ (w.l.o.g. in 
XLES-grid $1$):
\begin{align}
\label{eqn:PoorMansProjection}
    \lslt{{u}}{1,1}\left( \frac{-\Delta x_1}{2} + x_1\right) 
    &=    \lsltall{{u}}{1}^{\rm LES} \left(-\frac{\Delta 
    x_1}{2}\right)\\
      &- \int_{-\frac{\Delta x_1}{2}}^{-\frac{\Delta x_1 }{2} + x_1 } 
	\partial_{x_{2}} \lslt{{u}}{1,{2}} \dint{x'_1}
      - \int_{-\frac{\Delta x_1}{2}}^{-\frac{\Delta x_1}{2} + x_1} 
	\partial_{x_{3}} \lslt{{u}}{1,{3}} \dint{x'_1} 
\nonumber 
\end{align}
for $x_1\leq \Delta x_1 $. 
Owing to the absence of a 3D small scale velocity field, small 
scale pressure effects vanish from the equations: $\lslt{p}{} 
=\lsltall{p}{}^{\rm LES}$, but ODT 
explicitly models small time scale pressure effects (see section 
\ref{s:ODT}). 


Because 9 velocity components within 3 XLES-grids are redundant, one dynamical 
velocity component in each XLES-grid is omitted by multiplying a 
matrix of Kronecker deltas
\begin{align}
\label{eqn:DiracDelta}
 \mat{\ident}-\mat{\delta_{i}}  = 
  \begin{pmatrix}
     1-\delta_{1i} & 0 & 0 \\
     0 & 1-\delta_{2i} & 0 \\
     0 & 0 & 1-\delta_{3i} 
  \end{pmatrix}
  \; {\rm with} \,
  ( 1-\delta_{ki}) = \begin{cases}
                    0 , \, {\rm if} \, k = i \\
                    1 , \, {\rm else}
		   \end{cases}.
\end{align}
to the XLES momentum equation Eq. (\ref{eqn:XLES_PreFilMomentumGridN}), 
leading to
\begin{align}
 \label{eqn:XLES_PreFilMomentumGridN3}
  0 =& ( \mat{\ident}-\mat{\delta}_{i})\partial_{x_i}  
{\lsltall{\vec{p}}{}}^{\rm LES} +  \left( \partial_t  - \nu 
\sum_{j=1}^3 
\partial_{x_j}^2   \right) ( \mat{\ident}-\mat{\delta}_{i}) 
\vec{\lspace{u}{}}_{i} \\
+&( \mat{\ident}-\mat{\delta}_{i}) \sum_{j=1}^{3} 
\left(
\partial_{x_j}  
{\lslt{\vec{u}}{j}}* 
\vec{\lspace{u}{}}_{i} 
+  \mat{l}^{\dag}* \partial_{x_j}   \mat{C}
      \left(
		\lslt{\vec{u}}{j} 		* \lspace{\vec{u}}{i}
	      - \lsltall{\vec{u}}{j}^{\rm LES} 	* 
\lspaceall{\vec{u}}{i}^{\rm LES} \right) 
\right)
\nonumber \\
+& \mathcal{\vec{M}}_{\rm ODT}^{\delta} + \mat{l}^{\dag} * \mat{C}\, 
\mathcal{\vec{M}}_{\rm ODT}^{\delta} 
   + ( \mat{\ident}-\mat{\delta}_{i}) \vec{\sigma}{}  \nonumber .
\end{align}
Here terms to be modeled by the ODT model are
\begin{align}
 \label{eqn:ODTadvectionIdeal2}
  \mathcal{\vec{M}}_{ODT}^{\delta} = ( \mat{\ident}-\mat{\delta}_{i})
  \sum_{j=1}^{3} 
   \left( \mat{l}^{2D}
	 \mathcal{\vec{M}}_{ij}^{\rm XLES} + 
\partial_{x_i} \lsst{\vec{p}}{}
	+ \partial_{x_j}  {\lsst{\vec{u}}{j}}* 
\vec{\lspace{u}{}}_{i} 
   \right)
\end{align}
and involve two velocity components orthogonal to the highly 
resolved direction in each XLES-grid (introduced in section \ref{s:ODT}).

\section{From XLES to ODTLES}
\label{s:XLES2ODTLES}

In this section ODT is interpreted as a microscale model within the XLES 
approach.
In ODT a stochastic process mimics 3D turbulent advection within a 1D 
sub-domain. 
Thus, an interpretation of ODT in terms of the Navier-Stokes advection is not 
straightforward and ODTLES is not directly deducible from 
Navier-Stokes equations.

\begin{figure}
        \centering
	    \includegraphics[width=0.4\textwidth]{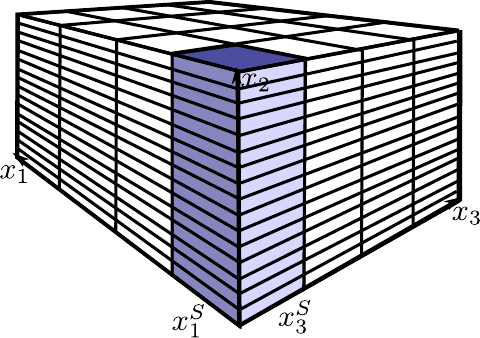}
	    \caption{Illustrative 1D stack in XLES-grid $2$ in $x_2$-direction 
at 
$(x_1^S,x_3^S)$}
	    \label{fig:ODTstack}
\end{figure}

Nevertheless XLES microscale terms w.l.o.g. in XLES-grid $2$ 
($\mathcal{M}_{2,ODT}^{\delta}$ in Eq. 
(\ref{eqn:ODTadvectionIdeal})) are 
interpreted by $N_{\rm LES_1} \times N_{\rm LES_3}$ so called stacks, each 
containing highly resolved 1D information, e.g. defined by one line at 
$(x_1^S, x_2,x_3^S)$ with constant $x_1^S$ and $x_3^S$ (see figure 
\ref{fig:ODTstack}). The microscale 
terms $\mathcal{M}_{2,ODT}^{\delta}(x_1^S,x_2, x_3^S)$ in each of these stacks 
$S$ contain $2$ velocity components and can be modeled by the ODT advancement 
(Eq. (\ref{eqn:eddyfunction})):
\begin{align}
   \mathcal{M}_{2,ODT}^{\delta}(x_1^S,x_2, x_3^S) = 
e_{2,i}(\lspace{u}{2,i};x_2,l) + \sigma_{ODT_k}  \, , 
\; {\rm and} \; l \leq l^{\rm max}=\Delta x^{\rm LES}_k
\end{align}
with the ODT model error $\sigma_{ODT_k}$. 
Note that the ODT model introduced in section \ref{s:ODT} also advances 
$2$ velocity components which are orthogonal to the $x_k$-direction. 

The maximum eddy length $l^{\rm max}$ within the ODT model corresponds to the 
largest (global) scale. 
By using ODT as a model within the XLES framework, the maximum eddy size 
$l^{\rm max}$ defines the boundary between turbulent 
scales described by the 3D advection scheme and by the ODT turbulent advection. 
Since ODT should capture turbulent effects not resolved by the 3D advection 
scheme, the maximum eddy size $l^{\rm max}$ mainly depends on the numerical 
properties of the 3D advection scheme and needs to be determined by numerical 
tests 
(not shown here). We found $l^{\rm max}=\Delta x^{\rm LES}_k$ to be convenient, 
which corresponds to the ability of the implemented numerical 3D advection 
scheme (see \cite[section 3.3]{Glawe:2014}) to resolve e.g. the Kolmogorov 
length scale 
with approximately one 3D cell in the `DNS-limit' of XLES (all scales are 
represented in 3D) as shown in part I (\cite[section 3.2]{Glawe:2014}). 

Closure of XLES can involve any form of modeling that specifies the RSS 
time
advancement on an entire XLES-grid such as XLES-grid 2 shown in figure 2.
This is not required to involve a collection of model instantiations on 
individual
stacks, such as the illustrative stack in that figure. Nevertheless, sub-grid 
ODT
within ODTLES is formulated in this way. On this basis, the XLES 3D advection
can be viewed as a form of coupling of the ODT instantiations within one grid. 
In
this context, the grid-to-grid coupling can be seen as a higher level of 
coupling.
This is mentioned because previous ODTLES formulations did not envision the
XLES framework with ODT not being the only conceivable RSS closure strategy
within an XLES grid

In previous ODTLES formulations, pressure projection was the only 
grid-to-grid
coupling. The insufficiency of this coupling is illustrated by a notional 
extension
of the method to include passive scalar properties. These are not subject to
pressure projection, so passive scalars on different grids would not be coupled 
at
all, an obviously unsatisfactory situation. The present approach is
straightforwardly extended to scalar properties in a manner that provides
appropriate grid-to-grid coupling. This conceptual advantage of XLES does not
imply that previous ODTLES formulations, which do not include scalars, are
necessarily deficient in some way. Indeed, for reasons that are not entirely 
clear
at present, they appear to perform well (\cite{RC-Schmidt2010}, 
\cite{ED-Gonzalez-Juez2011}), with the caveat that the latter documented 
oscillations in velocity root 
mean squares that are not seen in results obtained using the present 
formulation (see sections \ref{ss:ODTLESConvergence} and \ref{ss:ODTLESHighRe}).


\subsection{ODTLES: ODT-limit}
\label{s:ODT-Limit}

ODTLES includes another distinguished limit that we refer to as the 
`ODT-limit':
On the one hand ODTLES collapses to the ODT stand-alone model if only 
one 3D large scale cell represents the full domain. On the other hand the XLES 
microscale terms correspond to the full Navier-Stokes equations in this limit. 

Thus the ODT model error $\sigma_{\rm ODT}$ can be estimated by comparing ODT 
and 
DNS results for a turbulent channel (see \ref{app:ODTResults}).


The ability of ODT to describe the full spectrum of 3D turbulent effects is a 
required property to get an ODTLES 3D resolution largely independent of 
the turbulent intensity unless Reynolds-number variations trigger a global 
flow structure transition (see duct flow in section \ref{ss:ODTLESduct}).
Indeed, demonstrated model performance in the `ODT-limit' (i.e. ODT 
stand-alone, see \ref{app:ODTResults}) strongly indicates that ODT 
adequately describes the XLES model terms. This is also supported by ODTLES 
results that are shown to be in good agreement with DNS in sections 
\ref{ss:ODTLESConvergence}, \ref{ss:ODTLESHighRe}, and \ref{ss:ODTLESduct}.
Unfortunately a detailed theoretical investigation of the `ODT-limit' requires 
a 
convenient ODT interpretation in Navier-Stokes terms, which is not 
derived to a satisfying level yet, but ensemble 
statistics are formally analogous to corresponding Navier-Stokes terms to a 
considerable extent (e.g. the interpretation of ODT budget terms of the 
turbulent kinetic energy in \ref{app:ODTResults}).

\subsection{ODTLES: ODT Modeling Effects}
\label{ss:ODTeffects}

\begin{figure}
        \centering
        \begin{subfigure}[b]{0.49\textwidth}    
	  \includegraphics[width=\textwidth]{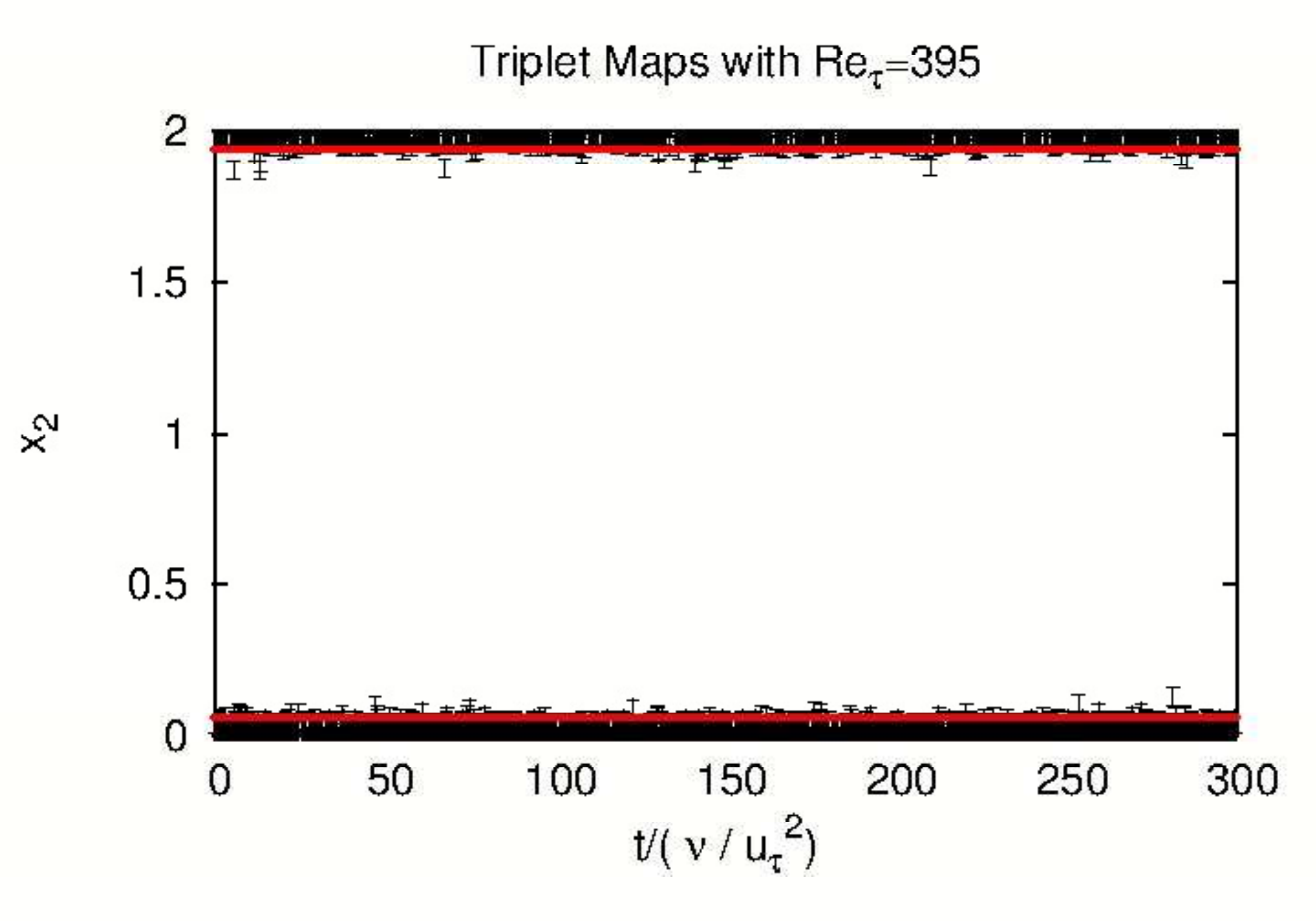}
                \caption{Eddies at $Re_{\tau}=395$.}
                \label{fig:Eddies395}
        \end{subfigure}     
        \begin{subfigure}[b]{0.49\textwidth}   
      \includegraphics[width=\textwidth]{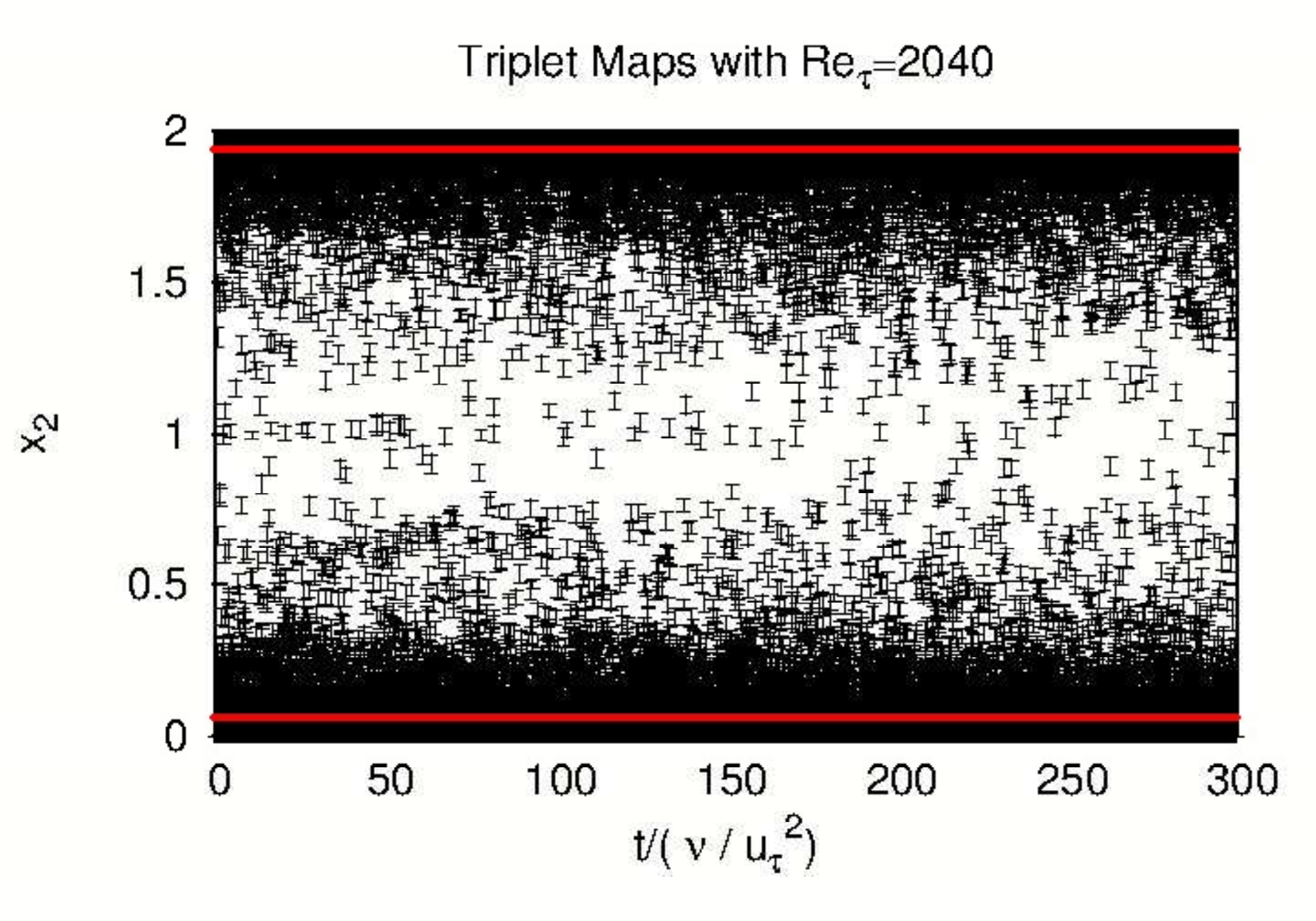}
                \caption{Eddies at $Re_{\tau}=2040$.}
                \label{fig:Eddies2040}
        \end{subfigure}         
        \caption{ODT turbulent events (`eddies') occurring in XLES-grid $2$ 
(superposition of eddies in all ODT domains). The size of the 3D cell closest 
to the walls is illustrated 
by red lines ($N_{\rm LES}=32$). For low Reynolds numbers, ODT primarily acts 
as 
a 
near-wall 
model (\ref{fig:Eddies395}); for high Reynolds numbers ODT acts as 
a sub-grid model over a larger extent of the flow domain 
(\ref{fig:Eddies2040}).}
\label{fig:Eddies}
\end{figure}

ODT introduces local turbulent events depending on the local flow state. 
In low Reynolds number channel flows the 3D grid is 
under-resolved only in the near-wall region (unless the grid is very coarse) and 
thus ODT works as a dynamical and highly accurate near-wall model, as figure 
\ref{fig:Eddies395} illustrates.
For highly turbulent flows, the 3D resolution in the core region of the channel 
 is under-resolved too: In this case ODT small scale eddy events 
 additionally occur in the core region 
introducing local turbulent transport effects, as figure 
\ref{fig:Eddies2040} illustrates. 

In contrast to wall-modeled LES, the ODT turbulent transport treats all regions 
consistently without introducing additional assumptions for the 
near-wall region. 

Additionally the ODT modeling depends on a fully resolved 
(1D) flow state and thus allows dominant small scale effects (e.g. local 
stratification, chemical processes, ...), which are not 
well captured by commonly applied eddy 
viscosity models.

\subsection{ODTLES: Coupled ODT Models}
\label{ss:CoupledODTmodels}

The turbulent ODT advection $e_{k,i}$ (in XLES-grid 
$k$) models the unresolved XLES terms $\mathcal{M}_{\rm ODT_k}^{\delta}$. These 
are  
coupled 
across the XLES-grids due to the SGS coupling term $\mat{l}^{\dag}* \mat{C}\, 
\mathcal{\vec{M}}_{\rm ODT}^{\delta}$ 
(see Eq. (\ref{eqn:XLES_PreFilMomentumGridN3})). 
The ODT diffusion terms $\mathcal{D}_{\rm ODT_k}=(-\nu) 
\partial_{x_k}^2 \lspace{u}{k,i}$ (for $k \neq i$) in XLES-grid $k$ 
represent the 
molecular diffusion 
as a continuum and are connected to the XLES diffusion terms 
$(\mat{\ident}-\mat{\delta}_i)(- 
\nu)\sum_{j=1}^{3} \partial_{x_j}^2 \lspace{\vec{u}}{k,i}$:
%
%
w.l.o.g. the XLES diffusion terms in grid $1$
\begin{align}	    
		  \nu  \partial_{x_1}^2 \lspace{u}{1,i}  
		+ \nu  \partial_{x_2}^2 \lspace{u}{1,i}  
		+ \nu  \partial_{x_3}^2 \lspace{u}{1,i}
	        \;, \; \, {\rm with} \, i=\{2,3\} 
\end{align}
are solved in three ways (compare 
to section \ref{ss:XLES_scales}):
\begin{enumerate}
 \item Diffusion directly resolved by ODT: $\mathcal{\vec{D}}_{\rm ODT}$

    The terms $-\nu  \partial_{x_1}^2 \lspace{u}{1,i}$ with $i=\{2,3\}$ are 
    interpreted by ODT incorporated in XLES-grid $1$. These terms are resolved 
    by $N_{\rm RSS}$ cells (representing molecular diffusion, 
similar to DNS)

 \item Diffusion indirectly resolved by ODT: $\mat{l}^{\dag}* \mat{C}\, 
\mathcal{\vec{D}}_{\rm ODT}$
 
    The terms $-\nu  \partial_{x_2}^2 \lspace{u}{1,3}$ and 
	      $-\nu  \partial_{x_3}^2 \lspace{u}{1,2}$ are
    interpreted by ODT domains residing in XLES-grid $2$ respectively XLES-grid 
$3$. 
    W.l.o.g. the first term is coupled from XLES-grid $2$ to XLES-grid $1$ by 
    $- [l_1^{-1}][l_2]  \nu  \partial_{x_2}^2 \lspace{u}{2,3}$
    (index notation for $\mat{l}^{\dag} * \mat{C}\, 
{\mathcal{\vec{D}}}_{\rm ODT}$). 
The diffusion is fully resolved, but additionally filtered (convolved) and 
deconvolved (see \cite[section 3.1]{Glawe:2014}). 
 
 \item Diffusion resolved by XLES: $\mathcal{\vec{D}}_{\rm XLES}$
 
    The terms $-\nu  \partial_{x_2}^2 \lspace{u}{1,2}$ and 
	      $-\nu  \partial_{x_3}^2 \lspace{u}{1,3}$ are not interpreted by 
ODT in any XLES-grid (this is caused by $(\mat{\ident} - 
\mat{\delta}_i)$; see section \ref{ss:XLES_mass}). A numerical 
interpretation is possible within XLES-grid $1$ using $N_{\rm LES}$ cells:
$-\nu  \partial_{x_i}^2 \lspace{u}{1,i}$ with  $i=\{2,3\}$.
 These diffusive terms are not resolved down to the molecular level. 
 The XLES resolved diffusion terms are written as $\mathcal{\vec{D}}_{\rm 
XLES}$ and numerically represented by an explicit 
Euler scheme in time and a spatial central difference scheme.

\end{enumerate}

In summary the ODT model is incorporated into XLES-grid $k$ by interpreting 
diffusive effects $\mathcal{D}_{\rm ODT_k}$ and the microscale advection terms 
$\mathcal{M}_{\rm ODT_k}^{\delta}$:
\begin{align}
  \mathcal{M}_{\rm ODT_k}^{\delta} - \nu \partial_{x_k}^2 \lspace{u}{k,i} 
\approx  
e_{k,i} + \mathcal{D}_{\rm ODT}(\lspace{u}{k,i}) , 
\,{\rm 
for} \,  
i \neq k
\end{align}
Additionally the (diffusive and advective) ODT terms are coupled between the 
XLES-grids by $\mat{l}^{\dag} * \mat{C}( \vec{e}_{i}(\lspace{\vec{u}}{i}) + 
\mathcal{\vec{D}}_{\rm ODT}(\lspace{\vec{u}}{i}))$. 
XLES diffusion terms $\mathcal{D}_{\rm XLES}$ are introduced to represent
diffusive terms not captured by ODT.

The under-resolved diffusion terms $\mathcal{\vec{D}}_{\rm 
XLES}$ are generally much smaller than the correct local diffusion and might be 
omitted in typical applications. Nevertheless these terms are conceptually 
desirable, because they allow the correct behavior in the `DNS-limit'.

\section{ODTLES: Time Advancement}
\label{ss:TimeAdvancement}

To advance the XLES equations in time a modified predictor-corrector procedure 
is used:
The XLES momentum equations including coupling terms and ODT advancement are 
solved, predicting velocity fields in each XLES-grid. 
Simultaneously time averaged 
velocity fields are computed.
A corrector step enforces the time averaged velocity fields to be 
divergence free (to ensure mass conservation).

Since the predictor-step involves ODT advancement and several coupling terms, a 
fractional time step algorithm is introduced: 
\begin{enumerate}
 \item predictor: 
    \begin{align}
    \label{eqn:timeIntegral}
     {{\lspace{\vec{u}}{i}}}^{*} =&  {{\lspace{\vec{u}}{i}}}(t)    +  
      \int_{t}^{t+\tau}  \partial_{x_i} {\lslt{\vec{p}}{}}_t \, 
    \dint t' \\
     \label{eqn:timeIntegral2}
     {{\lspace{\vec{u}}{i}}}^{**} =&  {{\lspace{\vec{u}}{i}}}^{*}   +  
    \int_{t}^{t+\tau} \sum_{j=1}^{3}  \partial_{x_j}     
      ({\lslt{\vec{u}}{j}}_t * \vec{\lspace{u}{}}_{i}^{*})  \, \dint t' \\
       \label{eqn:timeIntegral3}
     {{\lspace{\vec{u}}{i}}}^{***} =&  {{\lspace{\vec{u}}{i}}}^{**}   +
     \int_{t}^{t+\tau}  \mat{l}^{\dag} *\sum_{j=1}^{3} \partial_{x_j}
    \left(  {\lslt{\vec{u}}{j}}_t * \vec{\lspace{u}{}}_{i}^{*}
	  - {\lsltall{\vec{u}}{j}}^{\rm LES}_t * \vec{\lspaceall{u}{i}}^{*,\rm 
LES}     \right)   \, \dint t'  \\
     \label{eqn:timeIntegral5}
       {{\lspace{\vec{u}}{i}}}^{****} =&  
    {{\lspace{\vec{u}}{i}}}^{***}   +  
    \int_{t}^{t+\tau}  \left(e_i(\lspace{\vec{u}}{i}^{***})
      +\mathcal{{D}}_{\rm ODT} \right)(\lspace{\vec{u}}{i}^{***})\, \dint t'
    +\int_{t}^{t+\tau} \mathcal{{D}}_{\rm 
XLES}(\lspace{\vec{u}}{i}^{***})  
\, \dint t'\\
 \label{eqn:timeIntegral6}
      {{\lspace{\vec{u}}{i}}}(t+\tau) =&  
{{\lspace{\vec{u}}{i}}}^{****}
    +   \int_{t}^{t+\tau}  \mat{l}^{\dag}* \mat{C} \left( 
e_i(\lspace{\vec{u}}{i}^{***})
+\mathcal{{D}}_{\rm ODT} \right)(\lspace{\vec{u}}{i}^{***})  \, \dint t'
    \end{align}
 \item corrector:
    \begin{align}
      &{\rm 1D\, filter}&  \lsltall{\vec{u}}{i}^{\rm LES}_{t+\tau}=& 
    \mat{l}^{1D}\lslt{\vec{u}}{i}_{t+\tau} \\
    &{\rm solve}&   0=&\sum_{i=1}^{3} \partial_{x_i} \lslt{\vec{u}}{i}_{t+\tau} 
  \rightarrow 
      \partial_{x_i} \lsltall{\vec{p}}{}^{\rm LES}_{t+\tau} 
    \end{align}
    with $\lslt{{\vec{p}}}{} =       
    \begin{pmatrix}
         l_1^{-1} & 0 & 0 \\
         0 & l_2^{-1} & 0 \\
         0 & 0 & l_3^{-1} 
    \end{pmatrix}
    \lsltall{{\vec{p}}}{}^{\rm LES} $ (no 
small scale pressure 
field within XLES).
\end{enumerate}
The subscript (e.g. $\ltime{\,}_t$) introduced in time averaged properties 
indicates the time $t$ 
(averaged over the last time step) respective $t+\tau$ (averaged over the 
actual 
time 
step with the time step size $\tau$). 


In the ODTLES advancement cycle the ODT advancement is the most costly 
sub-process which leads to a highly parallelizable algorithm.

The concrete numerical implementation of the individual steps 
Eq. (\ref{eqn:timeIntegral}--Eq. (\ref{eqn:timeIntegral6}) is shown in 
\ref{app:NumImpl}.


\section{Application: Channel Case}
\label{s:Application}

\subsection{ODTLES: Convergence Study}
\label{ss:ODTLESConvergence}

To verify the ODTLES model, we conduct a numerical convergence study by 
computing a turbulent channel flow with friction Reynolds number 
$Re_{\tau}=395$.

In part I \cite[section 3.4]{Glawe:2014} XLES-U (unclosed XLES) and LES-U 
(under-resolved, unclosed LES) results are compared to DNS by \cite{KAM99} 
(online available: 
\cite{Kawamura:2013}):
both LES-U and XLES-U converge towards DNS with increasing 3D resolution, 
but only XLES-U is able to represent the laminar sublayer independent of the 3D 
resolution. (The numerical schemes for LES-U and XLES-U are identical).

\begin{figure}
        \centering
        \begin{subfigure}[b]{0.49\textwidth}
		\centering        
                \includegraphics[width=0.9\textwidth]
                {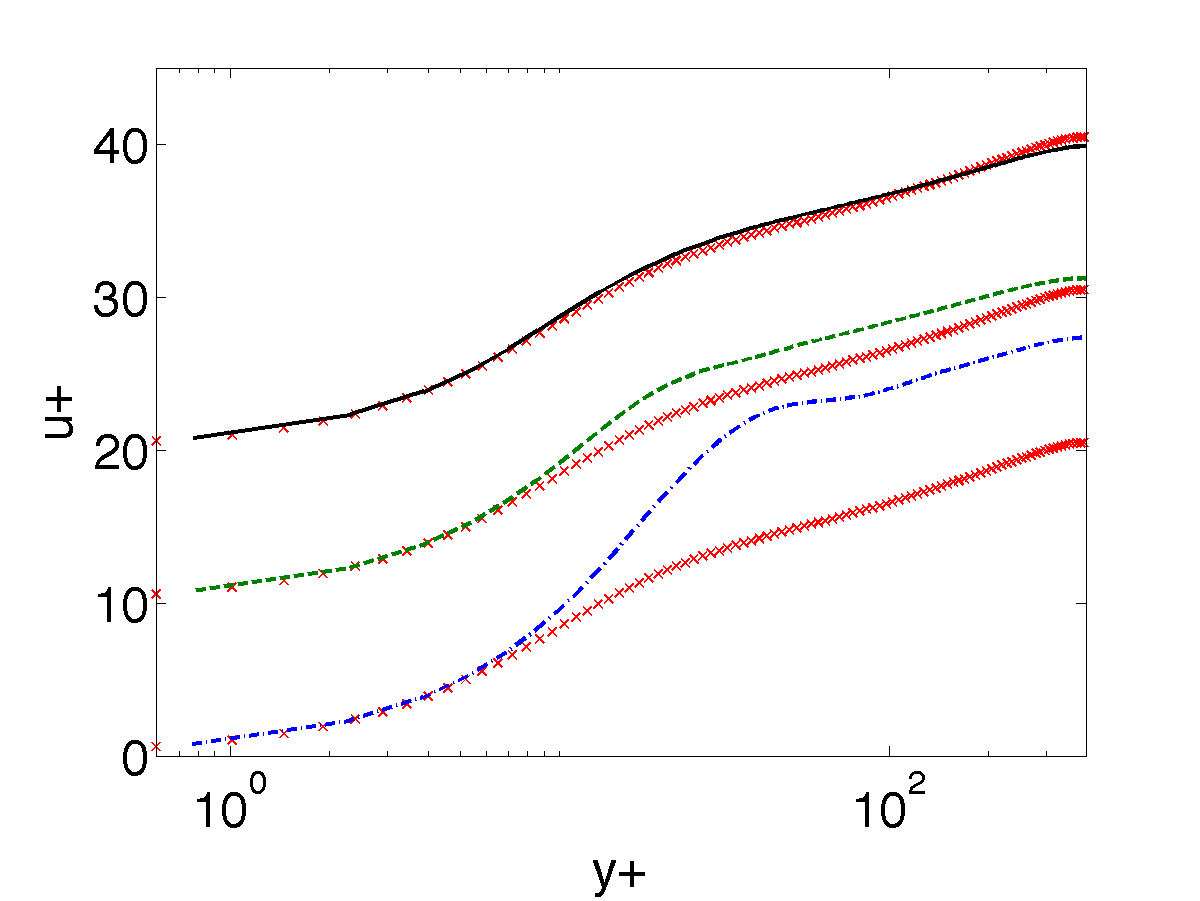}
                \caption{ XLES  : law of the wall ($\lspace{u}{2,1}$).  
Profiles shifted with increasing $N^{\rm LES}$.}
                \label{fig:umXLES}
        \end{subfigure}          
        \begin{subfigure}[b]{0.49\textwidth}
		\centering
		\includegraphics[width=0.9\textwidth]
		{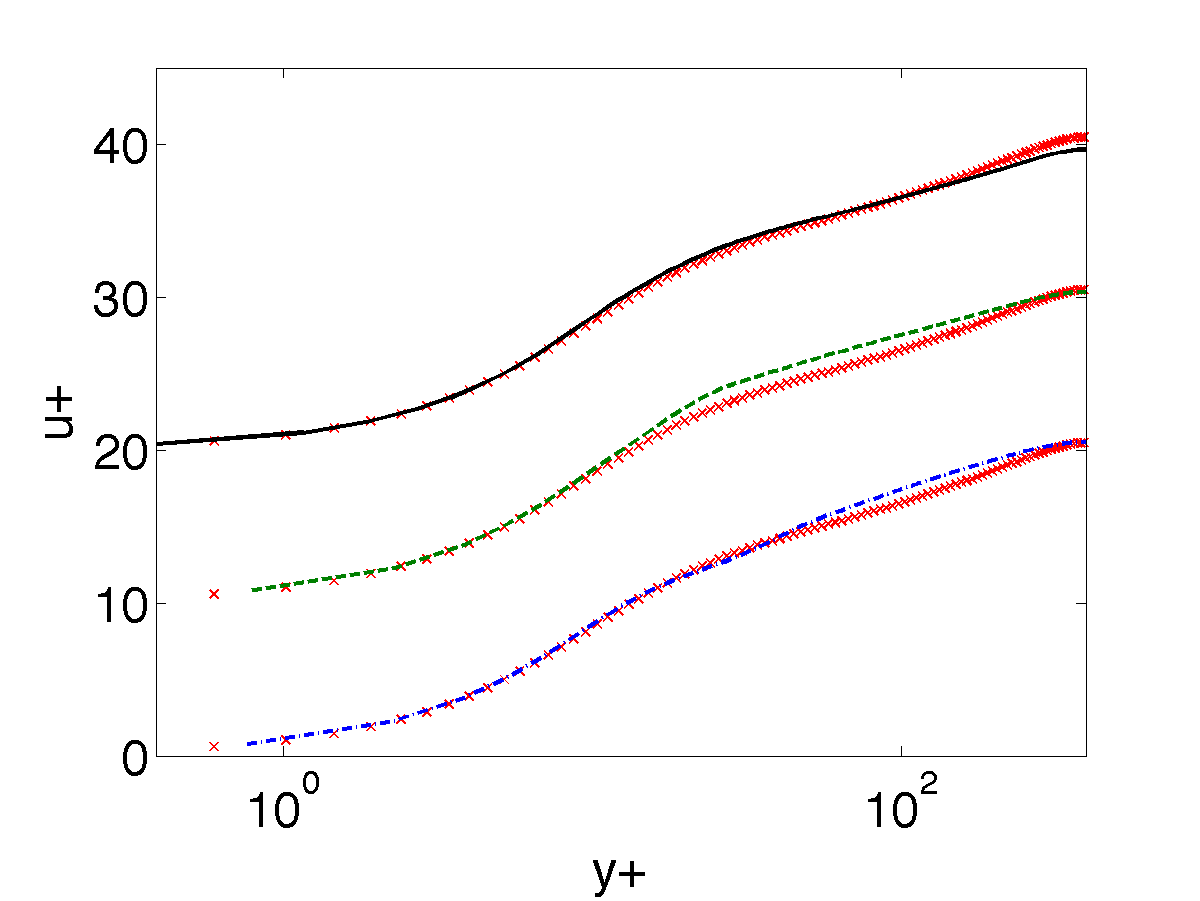}
                \caption{ ODTLES: law of the wall ($\lspace{u}{2,1}$). Profiles 
shifted with increasing $N^{\rm LES}$.}
                \label{fig:umODTLES}
        \end{subfigure}           
        \begin{subfigure}[b]{0.49\textwidth}  
	  \centering        
	  \includegraphics[width=0.9\textwidth]{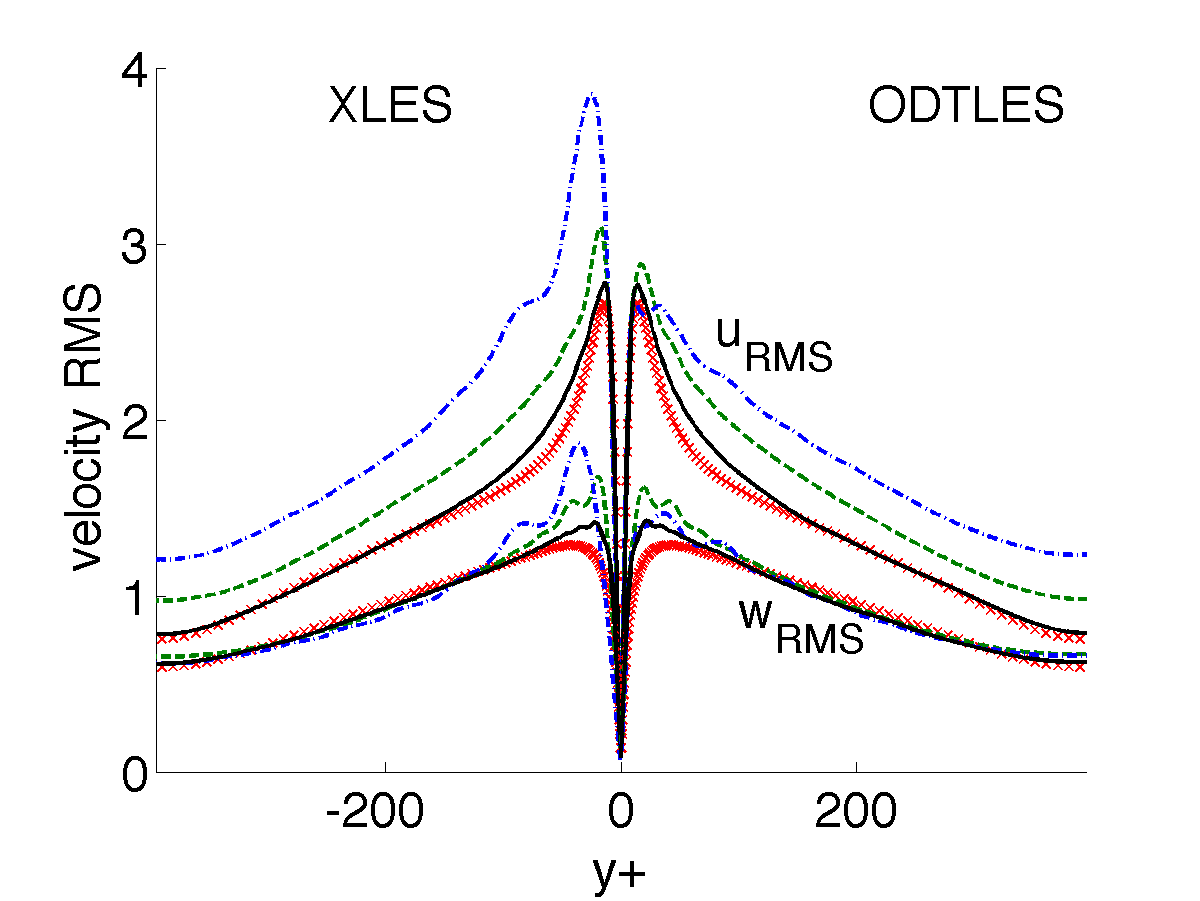}
                \caption{ Streamwise ($u_{\rm RMS}$) and spanwise  
($w_{\rm RMS}$) velocity RMS.}
                \label{fig:uvwRODTLES}
        \end{subfigure}
        \begin{subfigure}[b]{0.49\textwidth}
	  \centering               
                \includegraphics[width=0.9\textwidth]
                {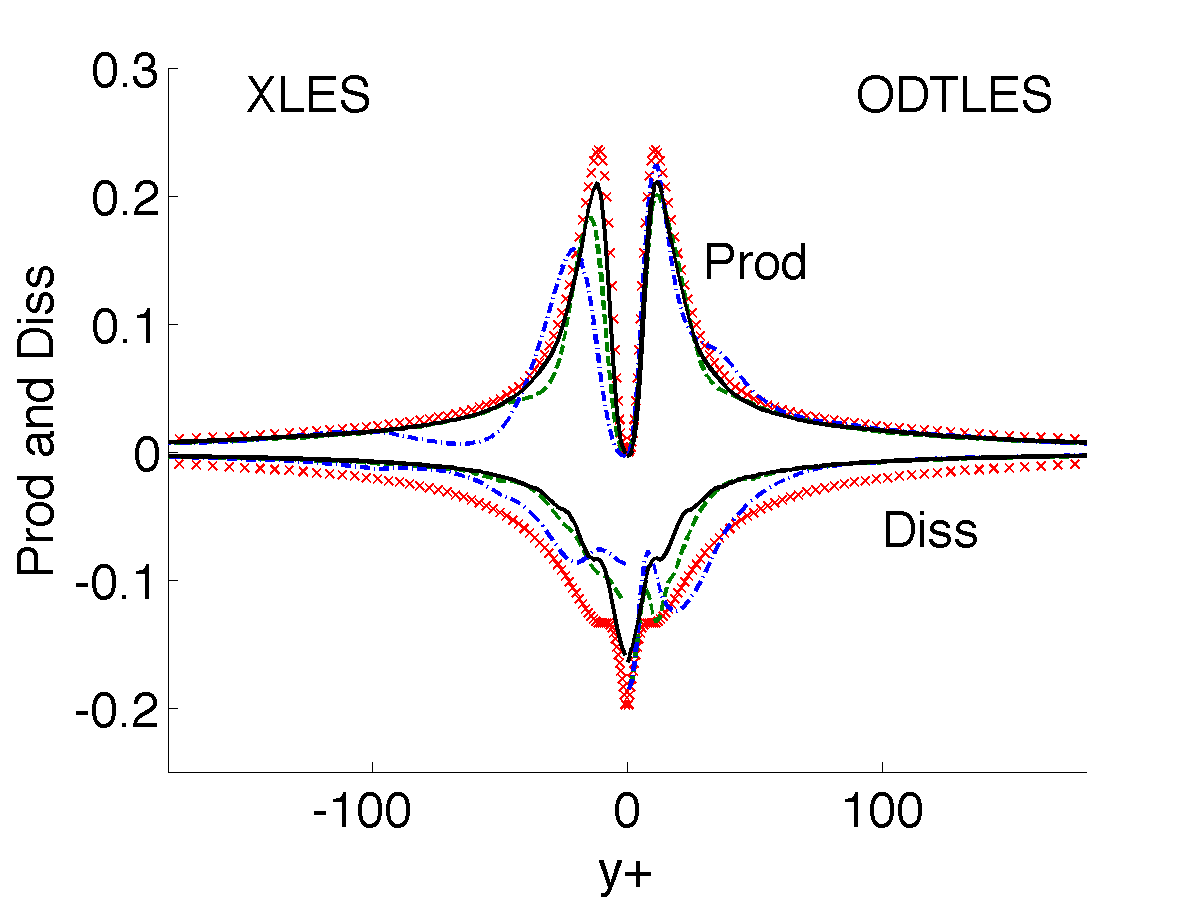}
                \caption{Production (Prod) and Dissipation (Diss) of the 
turbulent kinetic energy.}
                \label{fig:ProdDissODTLES}
        \end{subfigure}\\
        \begin{subfigure}[b]{0.49\textwidth}
	  \centering               
                \includegraphics[width=0.9\textwidth]
                {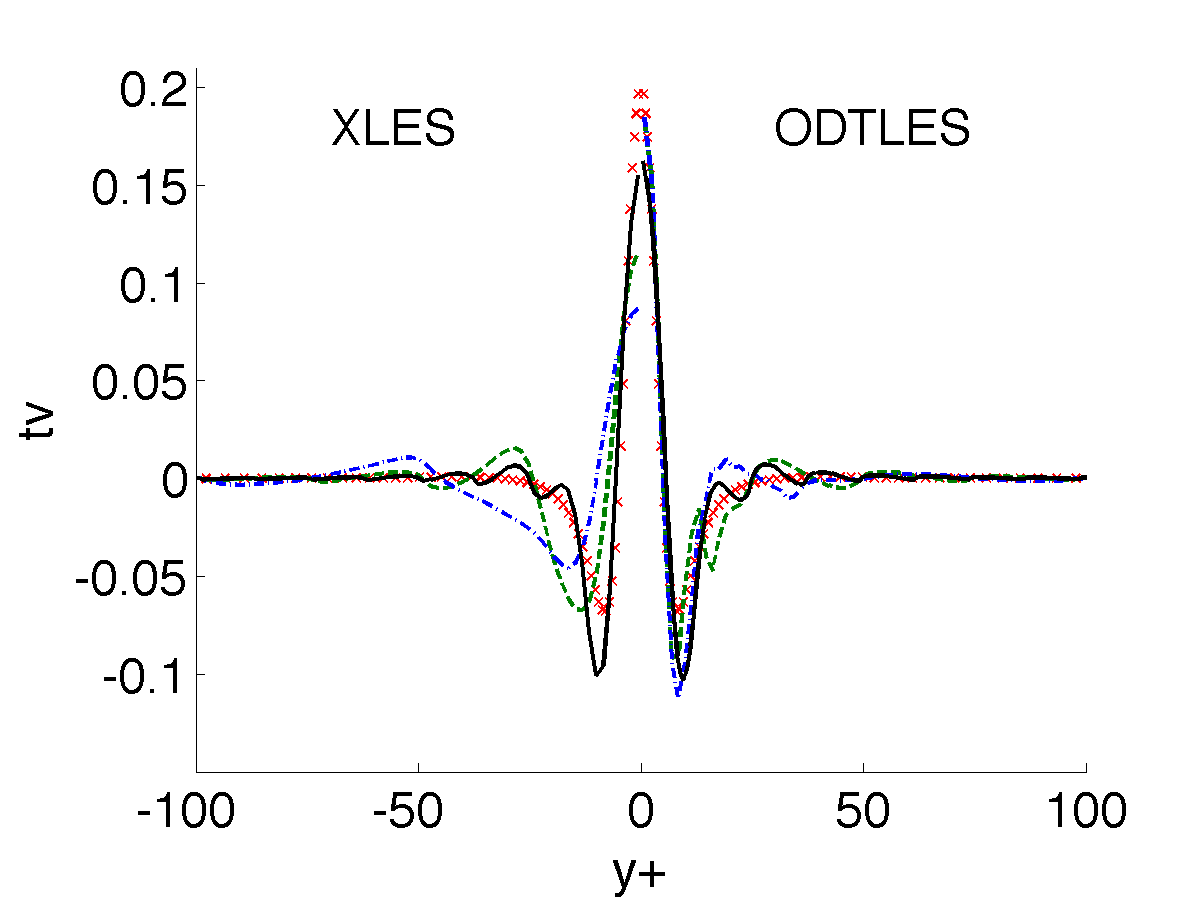}
                \caption{Viscous transport of the turbulent kinetic 
energy (tv).}
                \label{fig:TvODTLES}
        \end{subfigure}
        \begin{subfigure}[b]{0.49\textwidth}
	  \centering               
                \includegraphics[width=0.9\textwidth]
                {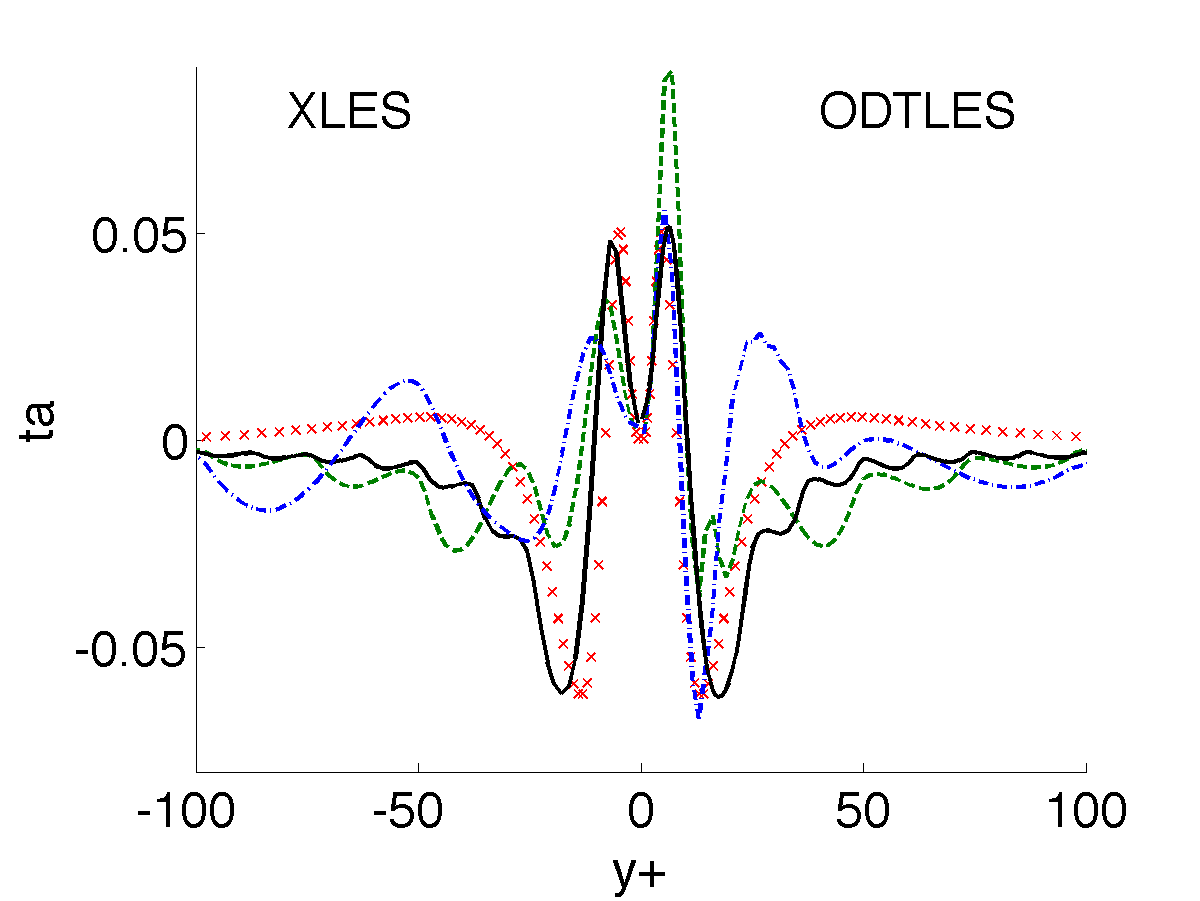}
                \caption{Advective transport of the turbulent 
kinetic 
energy (ta).}
                \label{fig:TaODTLES}
        \end{subfigure}        
	    \caption{Turbulent channel flow results for DNS 
(\textcolor{red}{small crosses}), ODTLES, and XLES-U with
$N_{\rm LES}=16$ 
(\textcolor{blue}{dash-dotted}), $N_{\rm LES}=32$ 
(\textcolor{darkgreen}{dashed}), $N_{\rm LES}=64$
(\textcolor{black}{solid}). The small scales are resolved using $N_{\rm 
RSS}=512$ cells. 
The flow statistics are based on the velocity field $\lspace{u}{2,i}$.}
	    \label{fig:Channel_UmODTLES}
\end{figure}

The ODTLES results are compared to XLES-U to show the 
significant effect of ODT as a SGM (see figure \ref{fig:Channel_UmODTLES}).
Additionally the DNS results are presented for comparison reasons.

For this convergence study, both XLES and ODTLES are using $N_{\rm 
LES}=\{16,32,64\}$ equidistant 3D large scale cells and $N_{\rm 
RSS}=512$  cells resolving the small scale. 

The XLES time step size is limited by Eq. (\ref{eqn:XLES_SmallTau})
 with the Courant-Friedrichs-Lewy number 
$CFL=0.45$ and the small scale cell size $\Delta x_{k,i}^{\rm RSS}$ (the ODT 
advancement is only indirectly influenced by this CFL condition). 
This choice minimizes the temporal XLES error terms 
$\Resttime{XLES}$ introduced due to time scale separation (see section 
\ref{s:TimeSeparation}).

The ODT model parameters are $C=6.5$ and $Z=330$ and match the ODT stand 
alone simulation presented in \ref{app:ODTResults}. The maximum eddy length 
$l^{\rm max}$ in XLES-grid $k$ equals the 3D large scale cell size $\Delta 
x_{k}^{\rm LES}$.

To produce statistically significant results, the flow is averaged for 
$t_{ave}> 25$ non-dimensional time units (compared to $t_{ave}=20$ for DNS) 
after  reaching a steady state.


The mean velocity profiles computed by ODTLES (see figure \ref{fig:umODTLES}) 
and XLES (see figure \ref{fig:umXLES}) are compared to DNS.

Additionally the streamwise and spanwise velocity RMS (see figure 
\ref{fig:uvwRODTLES}) and the budget terms of the turbulent kinetic energy (see 
figure \ref{fig:ProdDissODTLES}--\ref{fig:TaODTLES}) are shown.

As derived in \cite{AR-Kerstein2001} ODT pressure fluctuations cannot be 
distinguished from other turbulent transport terms. Thus we combine pressure 
terms into the turbulent transport also in ODTLES. 

XLES-U shows convergence towards the DNS results with 
increasing 3D resolution.
ODTLES is able to represent the flow field and turbulent statistics even with 
the very low 3D resolution $N_{\rm LES} = 16$, including the laminar sublayer 
near the walls and the budget terms of the turbulent kinetic energy. 
This significant improvement for channel flow results with very coarse 3D 
resolution compared to XLES-U suggests that ODTLES can encompass a wide 
Reynolds number range for fixed (coarse) 3D resolution.
With increasing 3D resolution ODT-specific issues in the near-wall statistics 
decrease in ODTLES.

\subsection{ODTLES: High Reynolds Number Flow}
\label{ss:ODTLESHighRe}

ODTLES combines the ability of the ODT model to describe all scales of 
highly turbulent flows within a 1D sub-domain of the full 3D domain with a 
coarse grained XLES approach representing the domain of e.g. a turbulent 
channel and 
introducing additional 3D effects compared to ODT stand-alone.

As shown in a convergence study in section \ref{ss:ODTLESConvergence} low 
Reynolds numbers are well described by ODTLES with only $N_{\rm LES}=16$ 3D 
cells.
To demonstrate the ODTLES ability to describe highly turbulent flows within a 
simple domain, we conduct turbulent channel flow computations with $N_{\rm 
LES}=32$ cells (in 3D) and up to $N_{\rm RSS}=16384$ cells to represent 
additional small scale effects, which allows Reynolds 
numbers $Re_{\tau}\leq 10000$.

The CFL number is chosen following Eq. (\ref{eqn:XLES_SmallTau}) with $CFL\leq 
1$ 
and ODT parameters are: $C=6.5$, $Z=330$, and $l^{\rm max}=\Delta 
x_k^{\rm LES}$. 

\begin{figure}
        \centering
        \begin{subfigure}[b]{0.49\textwidth}
		\centering
		\includegraphics[width=0.9\textwidth]
		{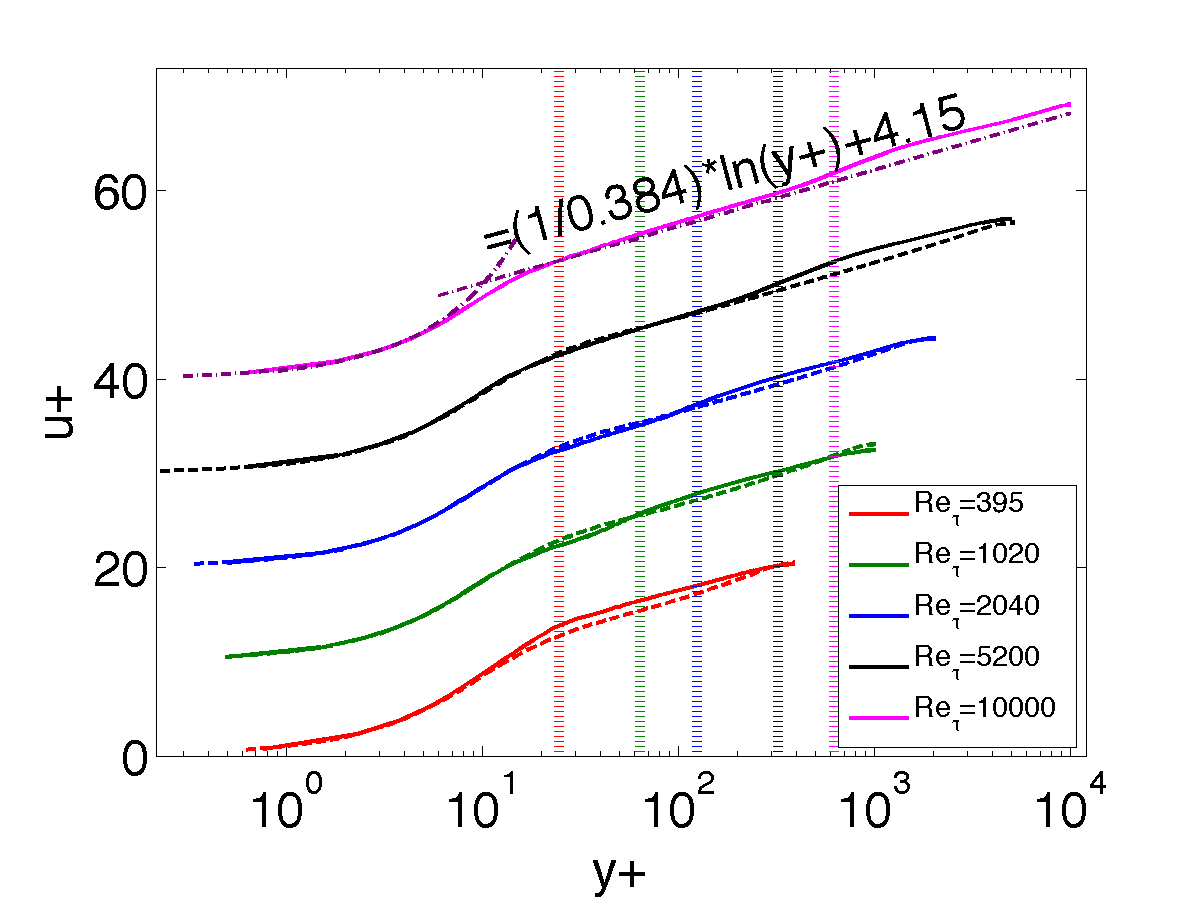}
                \caption{Law of the wall. Profiles shifted with increasing
$Re_{\tau}$.  }
                \label{fig:umODTLESX}
        \end{subfigure}            
        \begin{subfigure}[b]{0.49\textwidth}  
	  \centering        
	  \includegraphics[width=0.9\textwidth]
	  {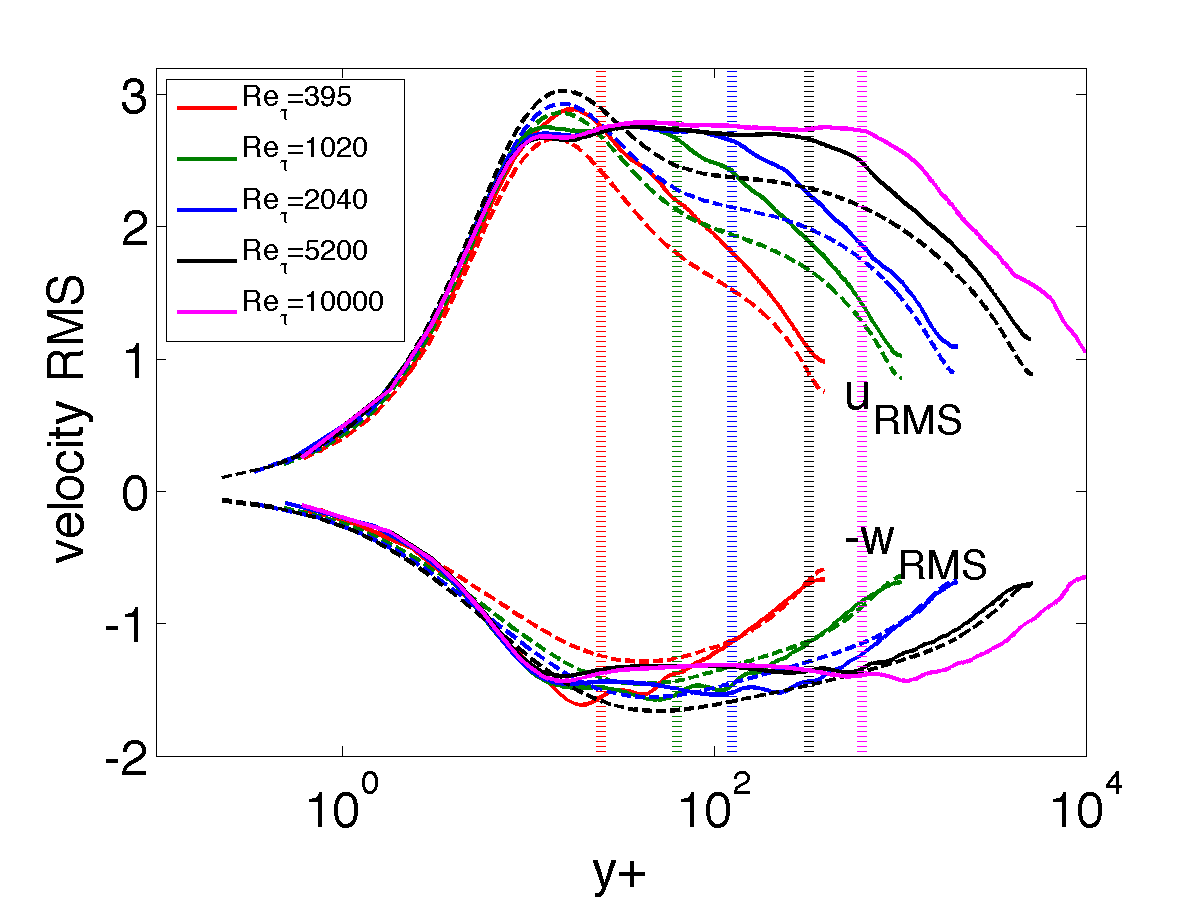}
                \caption{ Streamwise ($u_{\rm RMS}$) and spanwise  
($-w_{\rm RMS}$) velocity RMS.}
                \label{fig:uvwRODTLESX}
        \end{subfigure}
        \begin{subfigure}[b]{0.49\textwidth}
	  \centering               
                \includegraphics[width=0.9\textwidth]
                {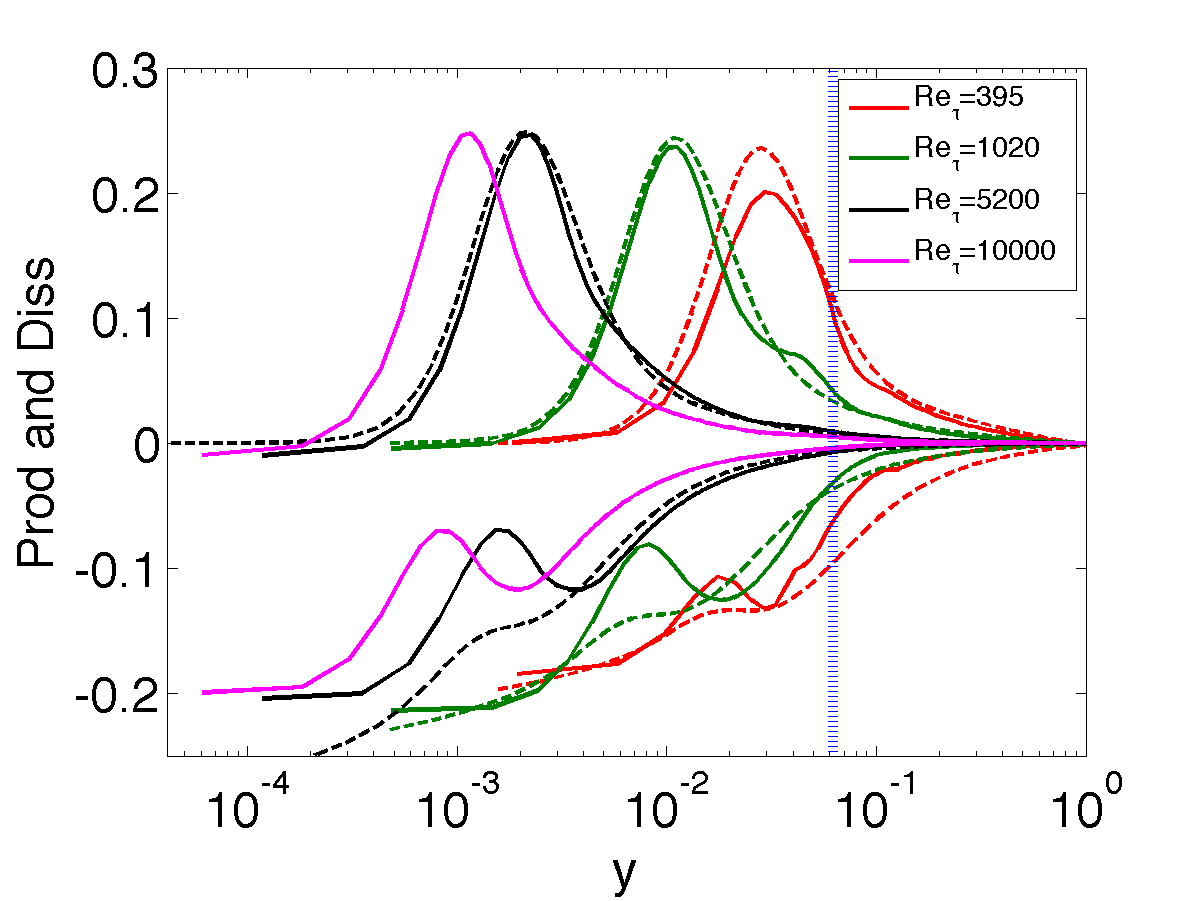}
                \caption{Production (Prod) and Dissipation (Diss) of the 
turbulent kinetic energy.}
                \label{fig:ProdDissODTLESX}
        \end{subfigure}
        \begin{subfigure}[b]{0.49\textwidth}
	  \centering               
                \includegraphics[width=0.9\textwidth]
                {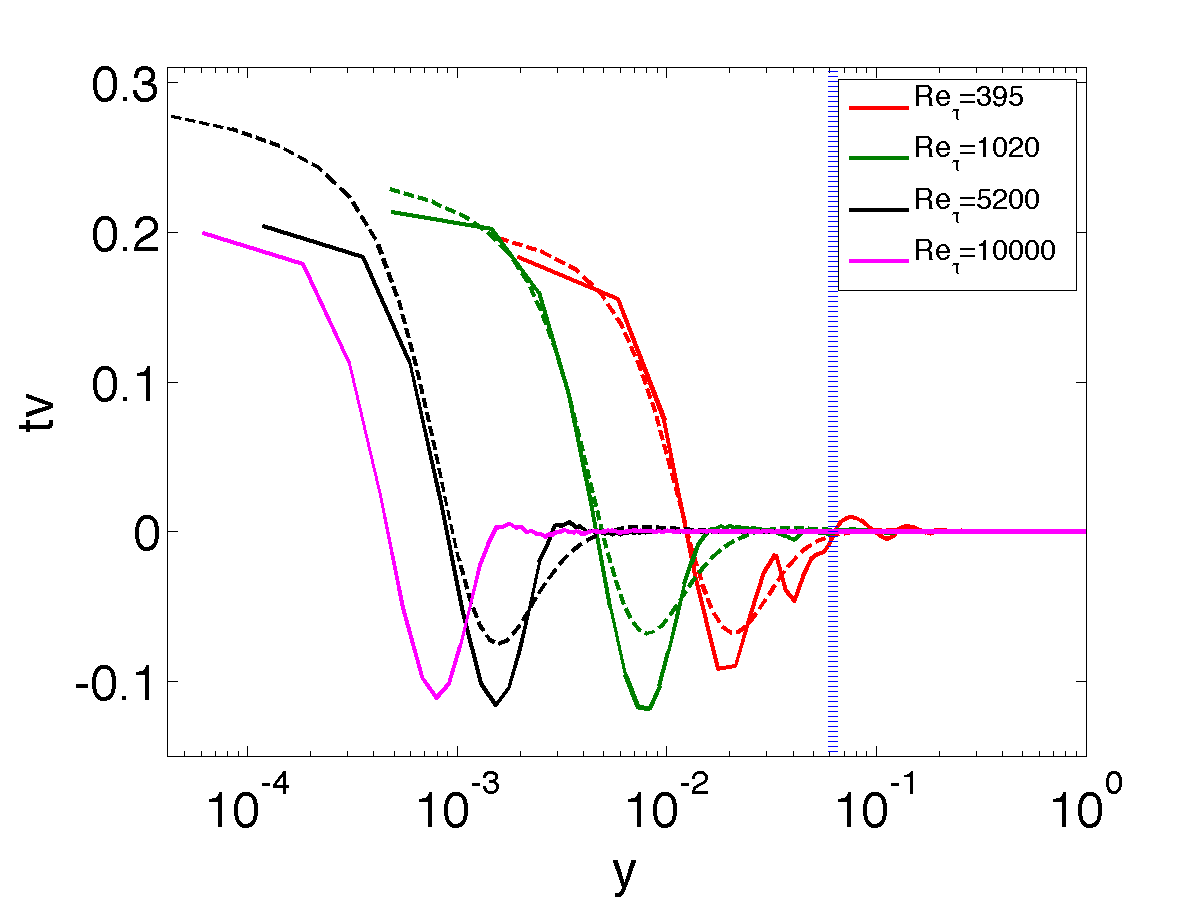}
                \caption{Viscous transport of the turbulent kinetic 
energy (tv).}
                \label{fig:TvODTLESX}
        \end{subfigure}
        \begin{subfigure}[b]{0.49\textwidth}
	  \centering               
                \includegraphics[width=0.9\textwidth]
                {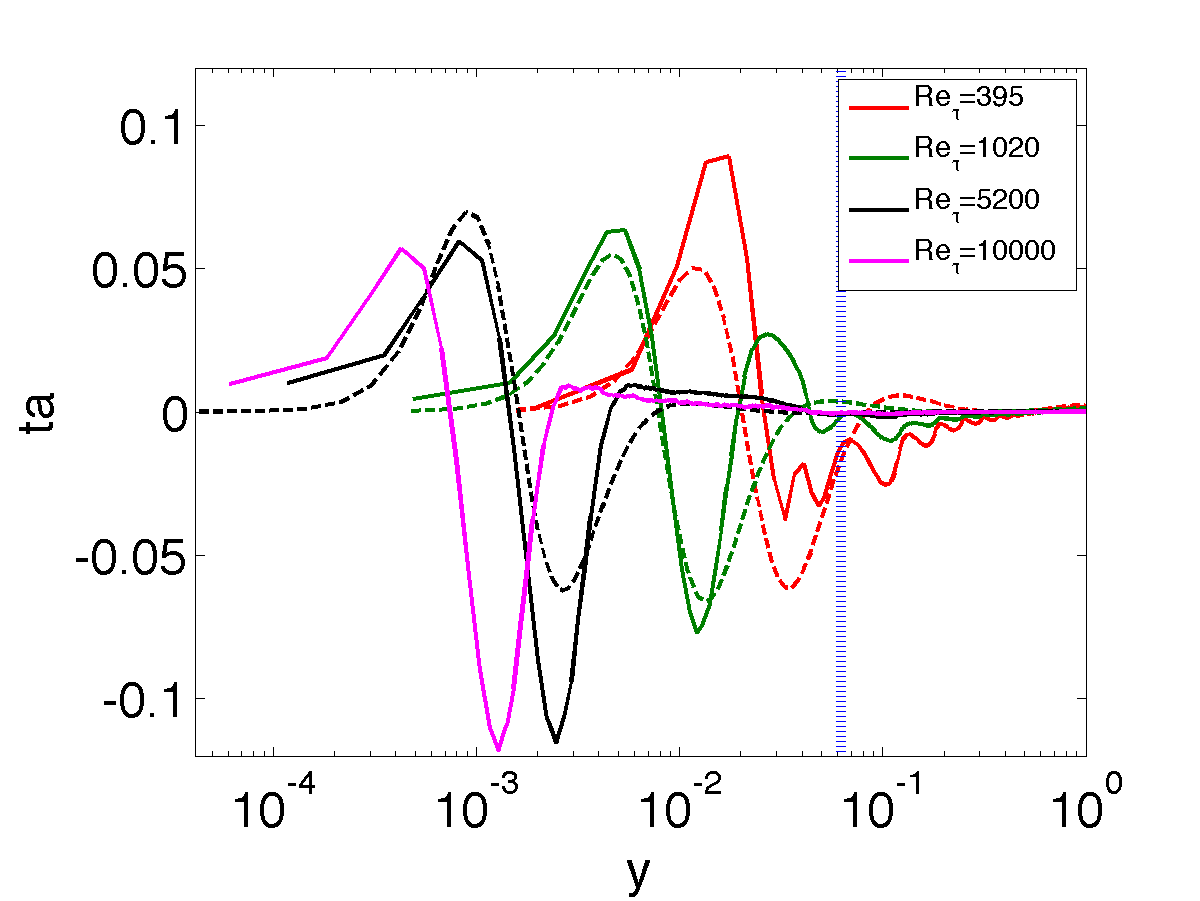}
                \caption{Advective transport of the turbulent kinetic 
energy (ta).}
                \label{fig:TaODTLESX}
        \end{subfigure}    
	\begin{subfigure}[b]{0.49\textwidth}
		\centering        
                \includegraphics[width=0.9\textwidth]
                {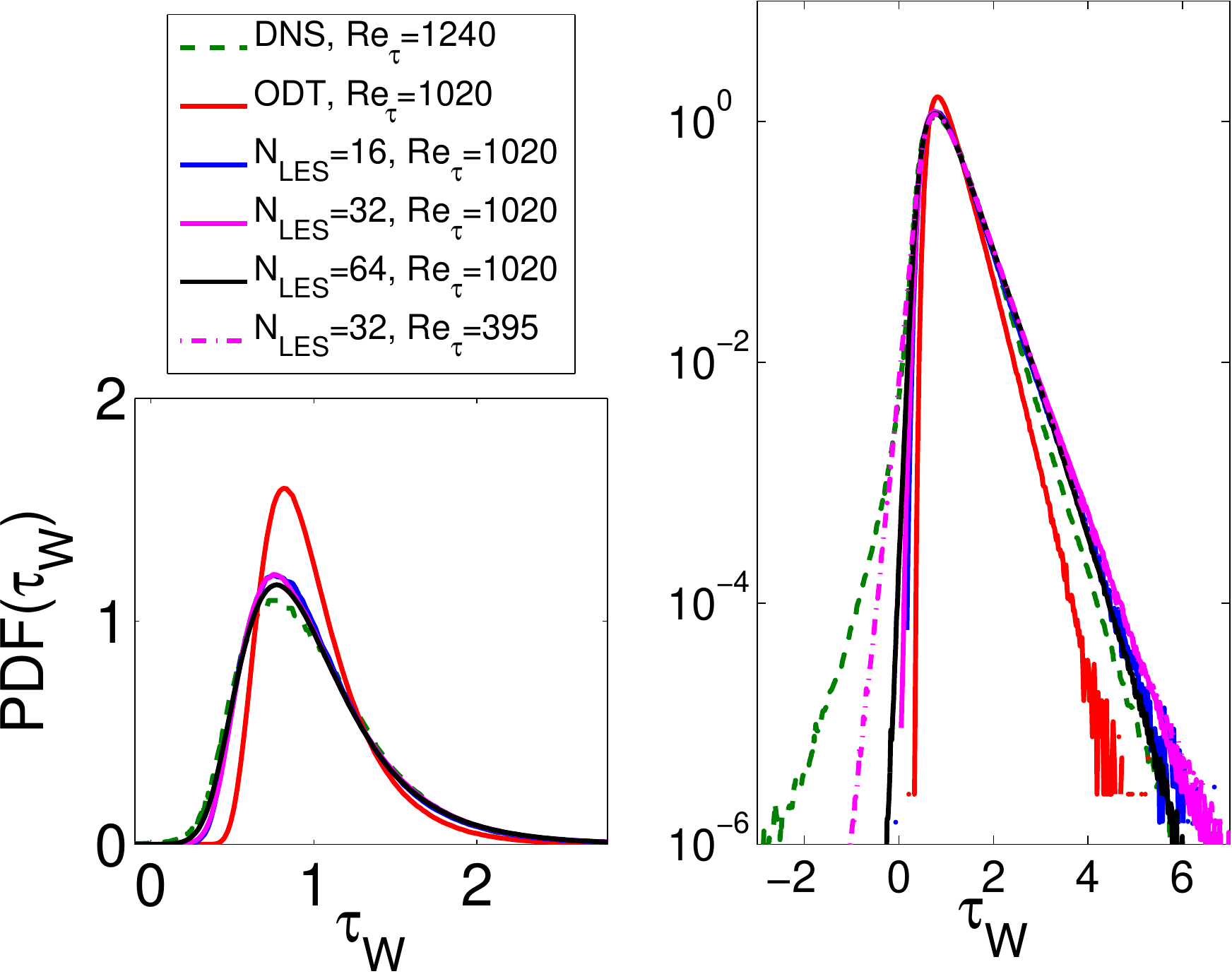}
                \caption{ Wall shear stress $\tau_W$: \\linear (left) and 
logarithmic (right). }
                \label{fig:TauWX}
        \end{subfigure} 
	    \caption{
	    Turbulent channel flow results for ODTLES (solid) and available DNS 
(dashed) for $N_{\rm LES}=32$ (if not indicated otherwise) and with 
$N_{\rm RSS}=512$  (for $Re_{\tau}=395$),  
$N_{\rm RSS}=2048$  (for $Re_{\tau}=1020$),
 $N_{\rm RSS}=4096$ (for $Re_{\tau}=2040$),
 $N_{\rm RSS}=8192$ (for $Re_{\tau}=5200$), and
  $N_{\rm RSS}=16483$ (for $Re_{\tau}=10000$). Figure \ref{fig:TauWX} compares 
the ODT and ODTLES wall shear stress ($Re_{\tau}=\{395,1020\}$) for various 
$N_{\rm LES}$ to DNS ($Re_{\tau}=1240$).
}
	    \label{fig:Channel_UmODTLESX} 
\end{figure}

The mean velocity profiles for several friction Reynolds numbers are 
illustrated in figure \ref{fig:umODTLESX} and compared to DNS 
by \cite{KAM99} and \cite{Moser:2014} (online available: 
\cite{Kawamura:2013} and \cite{Moser:2015}).
The ODTLES computation with $Re_{\tau}=10000$ is in good agreement with the 
laminar solution near the wall and the law of the wall with a von K\'arm\'an 
constant 
$\kappa = 0.384$, as obtained by \cite{Moser:2014} for $Re_{\tau}=5200$.

The streamwise and spanwise velocity RMS (see figure 
\ref{fig:uvwRODTLESX}) are in good agreement with the available DNS in the 
laminar region near the wall and beyond the first 3D cell. The size of the 
first 3D cell is illustrated by vertical lines for the different $Re_{\tau}$ 
values. 
Within the first 3D cell, ODT typically has some issues in representing the 
velocity RMS (this also 
applies for the budget terms of the kinetic energy).

In figure \ref{fig:ProdDissODTLESX}--\ref{fig:TaODTLESX} the budget terms of 
the turbulent kinetic energy are shown to be in good agreement with the DNS 
results, especially for highly turbulent flows.

ODT and ODTLES wall shear stress statistics for $Re_{\tau}=1020$, illustrated 
in 
figure \ref{fig:TauWX}, are compared to DNS results by 
\cite{Schlatter:2010} for $Re_{\tau} = 1240$. 
The ODT wall shear statistics are in rather good agreement with the DNS (see 
figure \ref{fig:TauWX} left), which indicates ODT to 
be an accurate near-wall model, yet by including additional 3D resolution 
$N_{\rm LES}$ within ODTLES the PDF is 
significantly improved.
A more detailed investigation (see figure \ref{fig:TauWX} right) shows 
that ODTLES 
underestimates rare backflow events. 
The reason could be that the responsible 3D structures near the wall are not 
represented due to the coarse 3D resolution, because the 3D cell size in 
wall units is $\Delta x^{\rm LES,+} \approx 32$ for the highest considered 3D 
resolution $N_{\rm LES}=64$ (with  $Re_{\tau} = 1020$). The result with lower 
Reynolds number ($Re_{\tau} = 395$) with 3D cell size in wall units $\Delta 
x^{\rm LES,+} \approx 24.7$ supports this hypothesis. Here we assume a low 
Reynolds number sensitivity of the wall shear stress statistics, as 
\cite{Schlatter:2010} report.

In summary ODTLES is able to capture the mean flow and turbulence statistics of 
highly turbulent flows up to $Re_{\tau}=10000$ within a simple domain. The 
computational costs are significant lower compared to DNS:
\cite{Lee:2013} report to use about $260 \times 10^6$ CPU hours on $786\times 
10^3$ cores for the DNS with $Re_{\tau}=5200$, while the corresponding ODTLES 
simulation takes $\approx 10000$ CPU hours on $24$ cores. 

A detailed investigation of the expected computational costs relative to 
RaNS, LES and DNS in section \ref{s:XLES_ResProps} shows ODTLES to be 
convenient to 
describe highly turbulent flows in domains of moderate complexity.

\subsection{ODTLES: Square Duct Flow}
\label{ss:ODTLESduct}

Sections \ref{ss:ODTLESConvergence} and \ref{ss:ODTLESHighRe} show turbulent 
channel flow results for the ODTLES model to be in good agreement with 
DNS. 
With increasing 3D resolution a convergence to DNS is observed, but even within 
an `ODT-limit' (corresponding to a single 3D cell) the ODT model reproduces key 
flow features stand-alone (see \ref{app:ODTResults}).

\begin{figure}
        \centering
	\begin{subfigure}[b]{0.49\textwidth}
		\centering
		\includegraphics[width=0.8\textwidth]
		{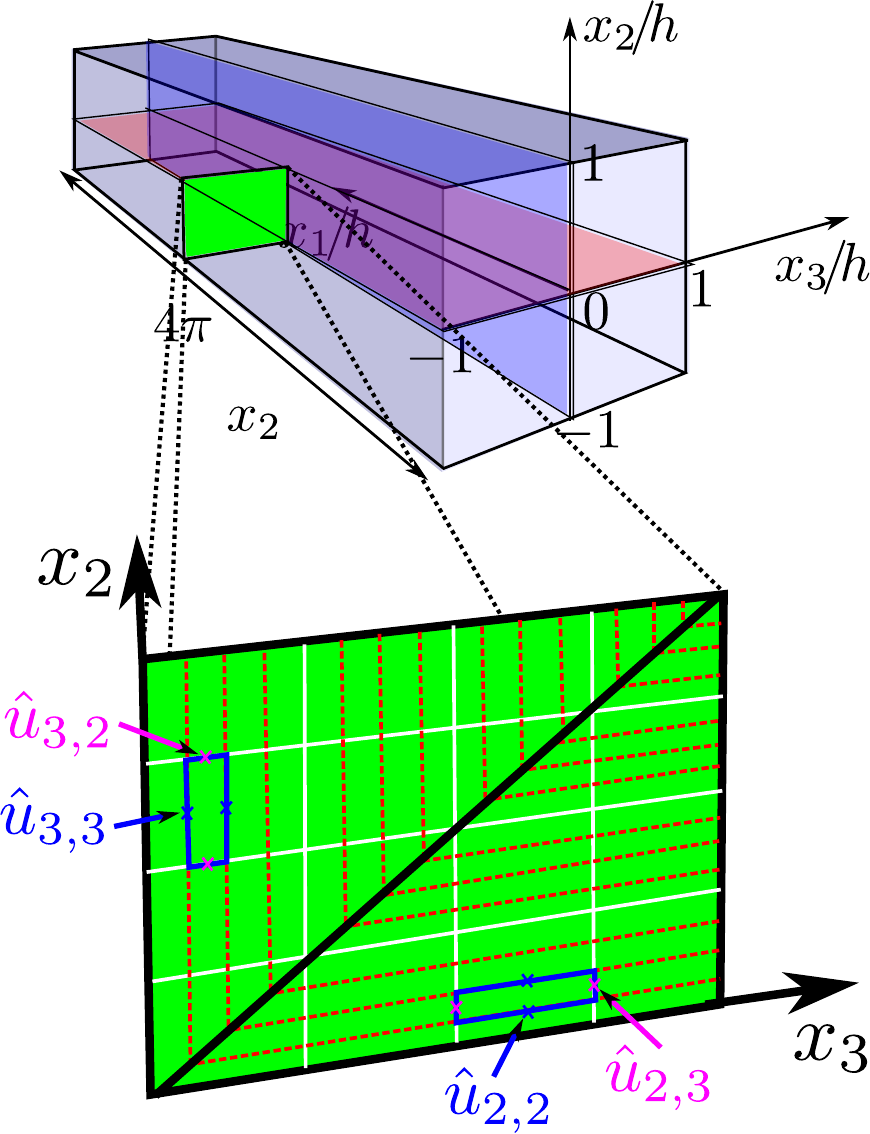}
                \caption{Geometry and illustrated XLES-grids.}
                \label{fig:geometryduct2}
        \end{subfigure}        
	\begin{subfigure}[b]{0.49\textwidth}
		\centering
		\includegraphics[width=0.8\textwidth]
		{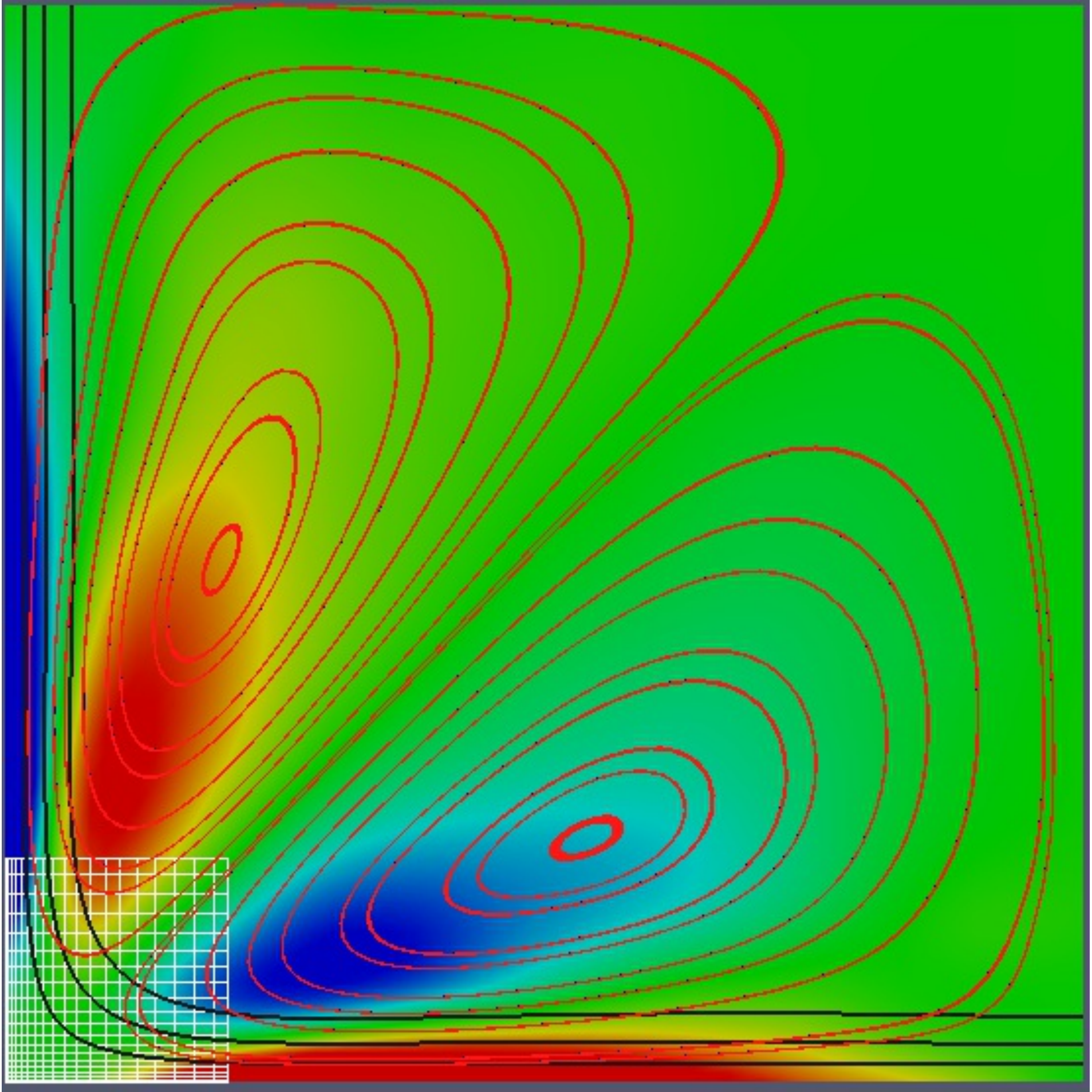}
                \caption{DNS:  ${Re_B}=2600$.}
                \label{fig:DuctDNSReB2600}
        \end{subfigure}         
        \begin{subfigure}[b]{0.49\textwidth}
		\centering
		\includegraphics[width=0.8\textwidth]
		{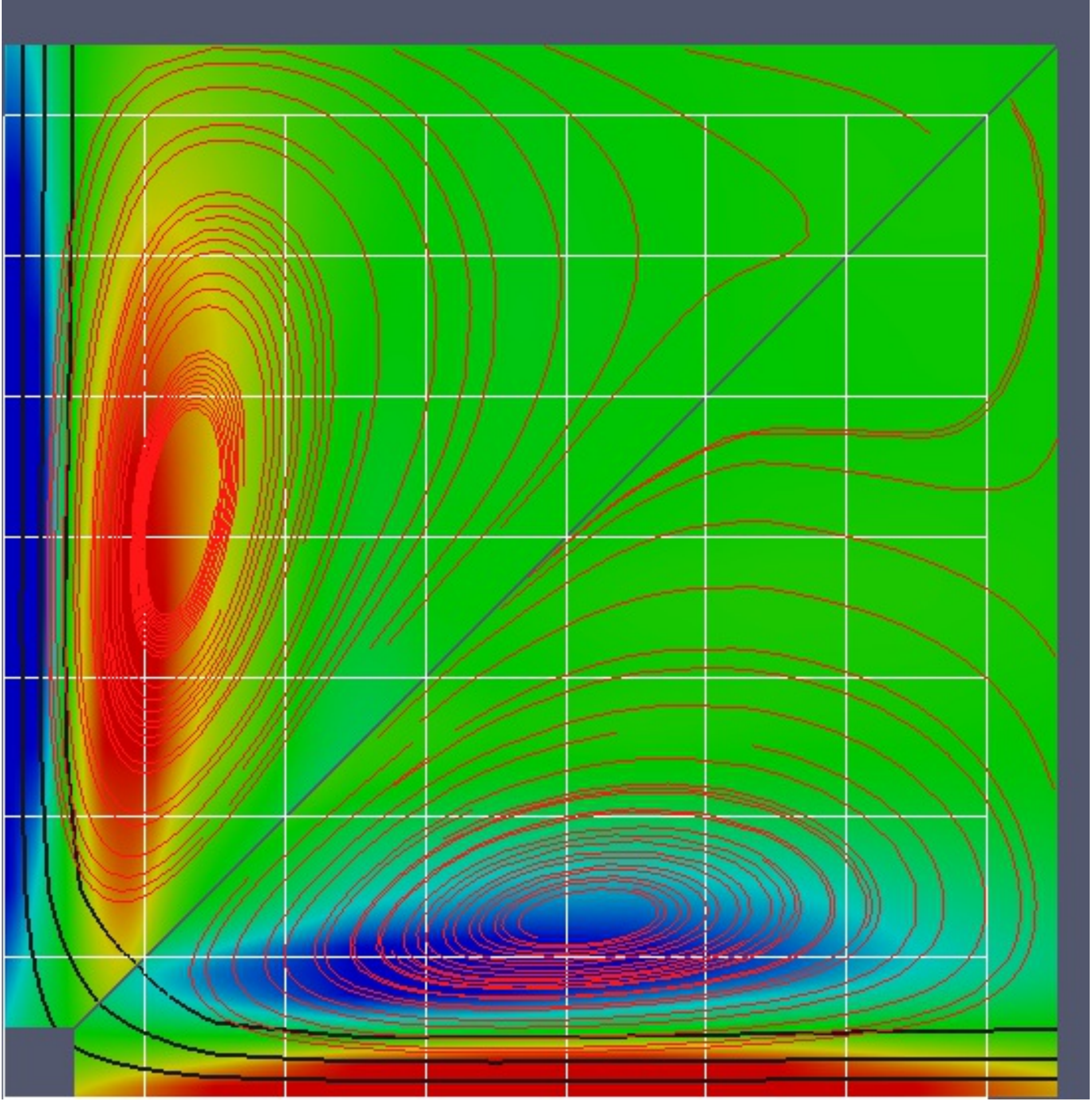}
                \caption{ODTLES:  $N_{\rm LES}=16$, ${Re_B}=2667$.}
                \label{fig:DuctODTLESN16ReB2600}
        \end{subfigure}  
        \begin{subfigure}[b]{0.49\textwidth}
		\centering
		\includegraphics[width=0.8\textwidth]
		{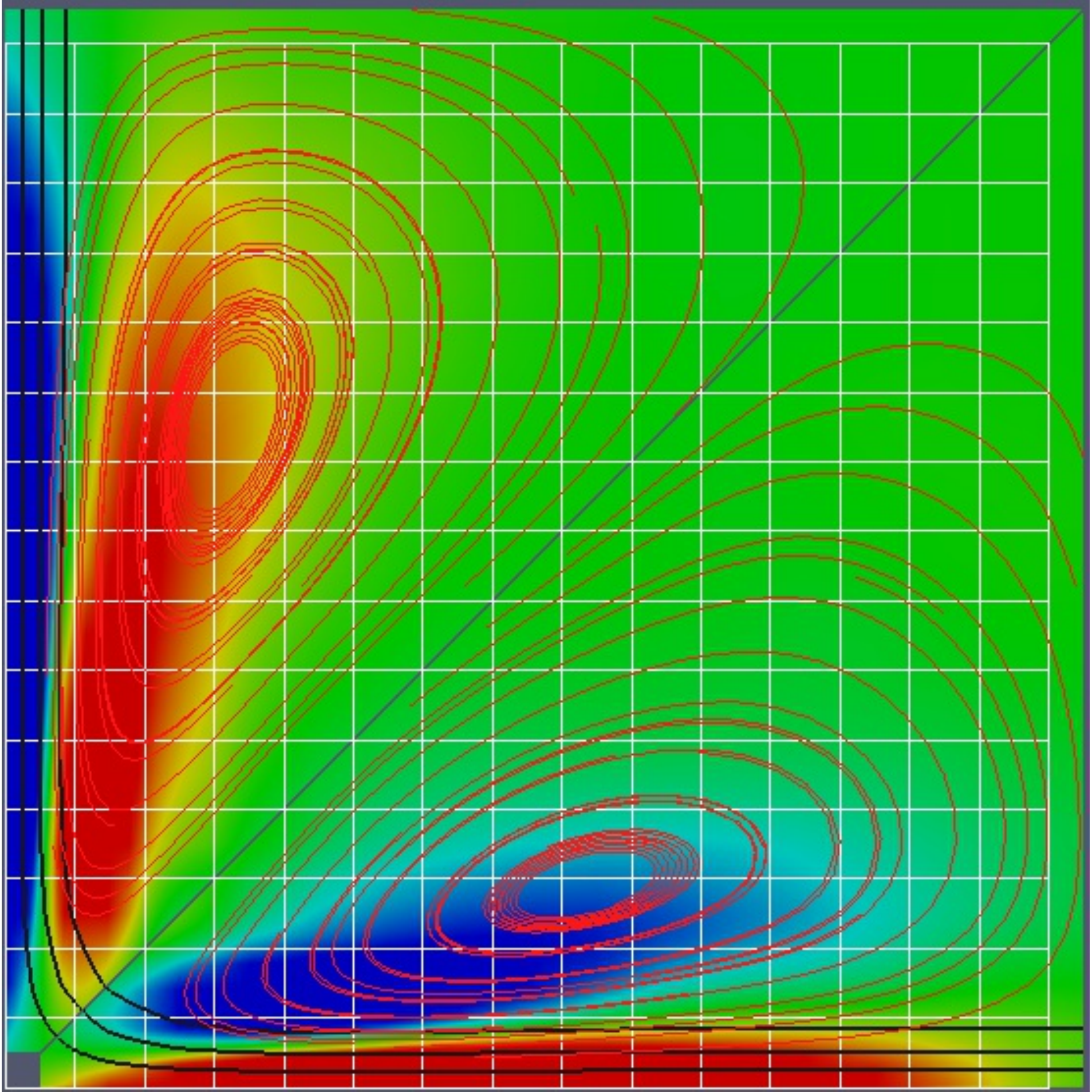}
                \caption{ODTLES:  $N_{\rm LES}=32$, ${Re_B}=2514$.}
                \label{fig:DuctODTESN32ReB2600}
        \end{subfigure} 
	    \caption{ All results are averaged in time and
streamwise direction. One quadrant of the duct is 
shown (see \ref{fig:geometryduct2}). 
The 3D grid is indicated by white lines (for the DNS only in the corner 
region). ODTLES properties are illustrated like cell centered and show 
additional small scale features (resolved by $N_{\rm RSS}=512$ cells) using the 
XLES-grid highly resolved in vertically $x_2$-direction (horizontally 
$x_3$-direction) in the lower right (upper left) triangular region, as 
illustrated in \ref{fig:geometryduct2}.
Contour lines of the primary mean flow (black) for 
$u_1=\{0.2,0.4,0.6\} \max(u_1)$, streamlines of the secondary mean 
flow ($u_2,u_3$) in red and the 2D vorticity $\omega_{2D}=\partial_{x_2}u_3 - 
\partial_{x_3}u_2$ (RGB color coded) are shown. } \label{fig:DuctReB2600} 
\end{figure}

In this section the square duct flow is investigated. This flow combines a 
simple geometry (see figure \ref{fig:geometryduct2}) and a complex flow 
behavior including secondary instabilities (secondary flow of Prandtl's second 
kind) and turbulent fluctuations. 
These secondary instabilities represent a 3D flow phenomena which is not
captured by the ODT model. For low Reynolds numbers (near the 
value for sustained turbulence) the size of the secondary flow structures in 
cross-stream direction corresponds to the half duct 
height ($h$). This flow regime in a square duct is investigated by 
\cite{Uhlmann:2007} using DNS.
Even for higher Reynolds number the cross-stream extension of the 
secondary instabilities is rather large scale compared to turbulent 
fluctuations occurring e.g. near the wall.
Nevertheless these small scale fluctuations play an important role for the 
duct flow because they generate secondary instabilities. 
The ODT model was shown to accurately describe small scale 
fluctuations and the secondary instabilities correspond to a rather large 
scale 3D flow feature which can be described by the XLES framework. Thus ODTLES 
is a highly promising model to describe the duct flow behavior, as 
\cite{ED-Gonzalez-Juez2011} and \cite{Glawe2013} showed in previous 
ODTLES studies. In these works no coupling terms (last line in Eq. 
(\ref{eqn:XLES_PreFilMomentumGridN})) between the XLES-grids are considered. 

The CFL number is chosen following Eq. (\ref{eqn:XLES_SmallTau}) with $CFL\leq 
1$.
The ODT model parameters are $C=6.5$, $Z=330$, and $l^{\rm max}=\Delta 
x_k^{\rm LES}$, which is identical to the channel flow setup in section 
\ref{ss:ODTLESHighRe}. 

The flow is averaged for $t_{ave} u_{B}/h \geq 2800$ non-dimensional time units 
(with the bulk velocity $u_B$) after reaching a steady state which is assumed 
to be sufficient to investigate the secondary instabilities. This is supported 
by a study of vortex structures by \cite{Uhlmann:2007} where nearly symmetric 
$8$-vortex structures are observed for lower averaging times at the investigated 
Reynolds numbers $Re_B \geq 2600$. Additionally the ODTLES results are averaged 
over the $4$ quadrants. 
Note that the DNS by \cite{Pinelli2010} uses $t_{ave} u_{B}/h \geq 7000$ 
to produce meaningful high order statistics.

Figure \ref{fig:DuctReB2600} compares the secondary flow computed by ODTLES for 
a moderate bulk Reynolds number $Re_B \approx 2600$ with the DNS by 
\cite{Pinelli2010} (online available: \cite{Uhlmann:2013}). Hereby ODTLES uses 
$N_{\rm LES}=\{16,32\}$ 3D large scale cells per direction and the
XLES specific small scale properties are resolved by $N_{\rm RSS}=512$ cells. 

The primary flow and key features of the secondary flow are in good agreement 
with the DNS results even with the very low 3D resolution of $N_{\rm LES}=16$ 
cells. Furthermore the ODTLES results indicate a convergence towards the DNS 
results with increasing 3D resolution. 

\begin{figure}
        \centering
        \begin{subfigure}[b]{0.49\textwidth}
		\centering
		\includegraphics[width=0.7\textwidth]
		{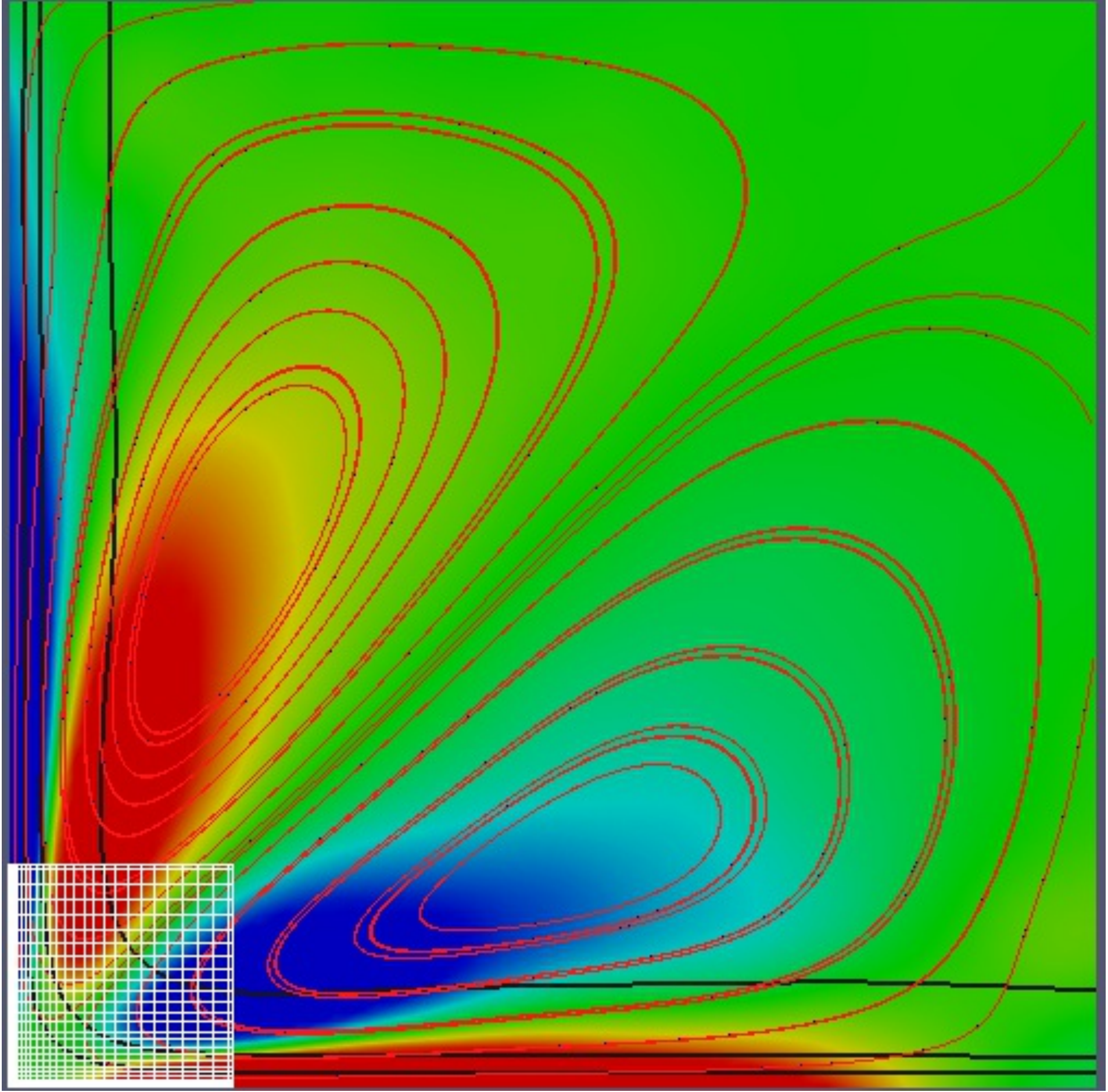} 
                \caption{DNS:\newline $Re_B=3500$.}
                \label{fig:DuctODTLESN32ReB2500}
        \end{subfigure}   
        \begin{subfigure}[b]{0.49\textwidth}
		\centering
		\includegraphics[width=0.7\textwidth]
		{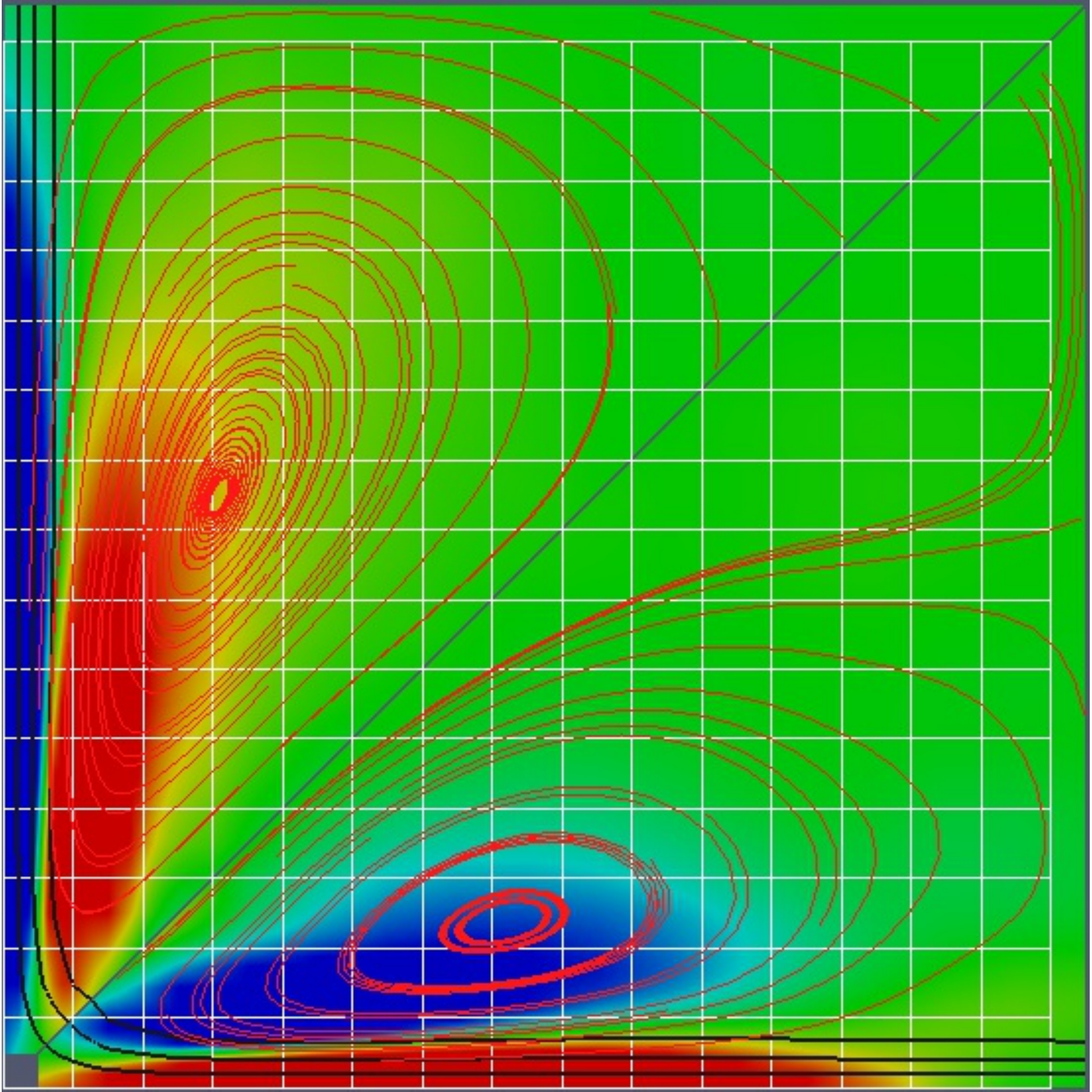} 
                \caption{ODTLES:\newline $Re_B=3485$, $N_{\rm RSS}=1024$.}
                \label{fig:DuctODTLESN32ReB3500}
        \end{subfigure}   
        \begin{subfigure}[b]{0.49\textwidth}
		\centering
		\includegraphics[width=0.7\textwidth]
		{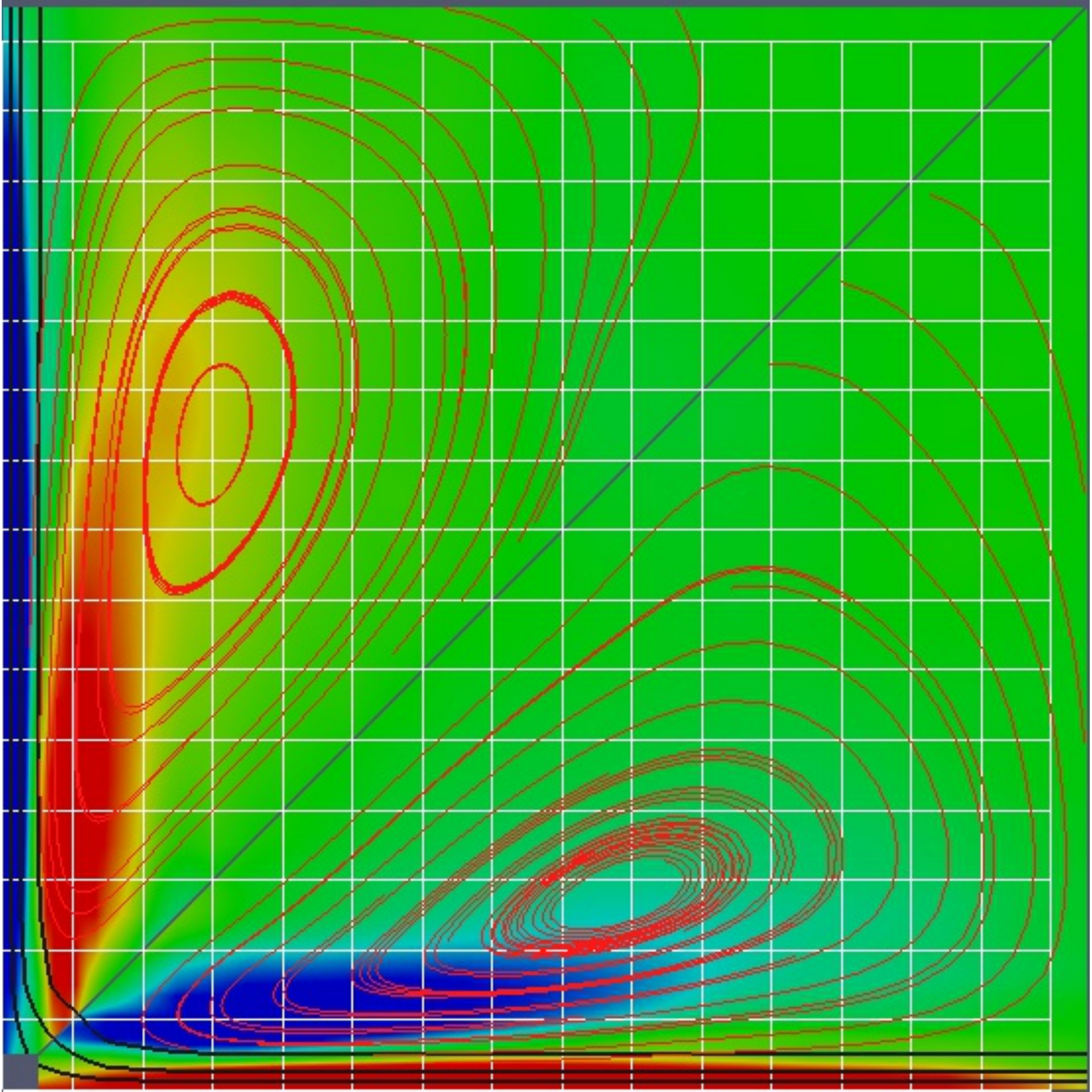} 
                \caption{ODTLES:\newline ${Re_B}=8446$, $N_{\rm RSS}=2048$.}
                \label{fig:DuctODTLESN32ReB8000}
        \end{subfigure}   
        \begin{subfigure}[b]{0.49\textwidth}
		\centering
		\includegraphics[width=0.7\textwidth]
		{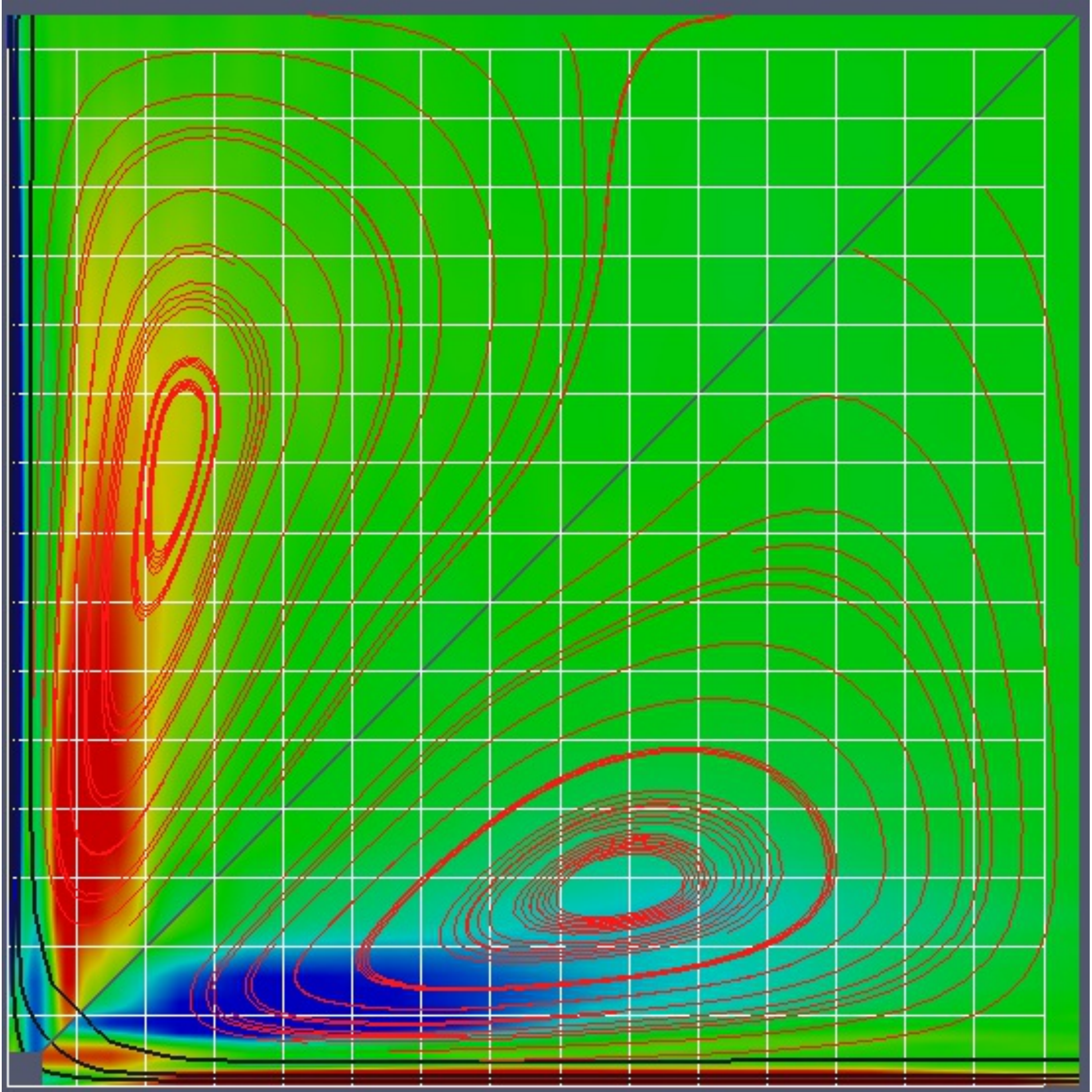} 
                \caption{ODTLES:\newline $Re_B=17338$, $N_{\rm RSS}=2048$.}
                \label{fig:DuctODTLESN32ReB17000}
        \end{subfigure}             
	    \caption{ 
 ODTLES and DNS square duct results with $N_{\rm LES}=32$ for different 
Reynolds numbers $Re_B$. Primary and secondary streamlines and vorticity are
illustrated similar to figure \ref{fig:DuctReB2600}. 
} 
\label{fig:DuctHighReynolds} 
\end{figure}

The secondary mean velocity field alters with increasing Reynolds number, 
as investigated in figure \ref{fig:DuctHighReynolds}. In particular each 
secondary vortex structure tends towards a triangular shape for increasing 
Reynolds number. The vorticity approaches the duct corner with increasing 
Reynolds number, indicating fast changes in secondary flow directions in this 
area.

\begin{figure}
      \centering
        \begin{subfigure}[b]{0.49\textwidth}
		\centering
		\includegraphics[width=0.7\textwidth]
		{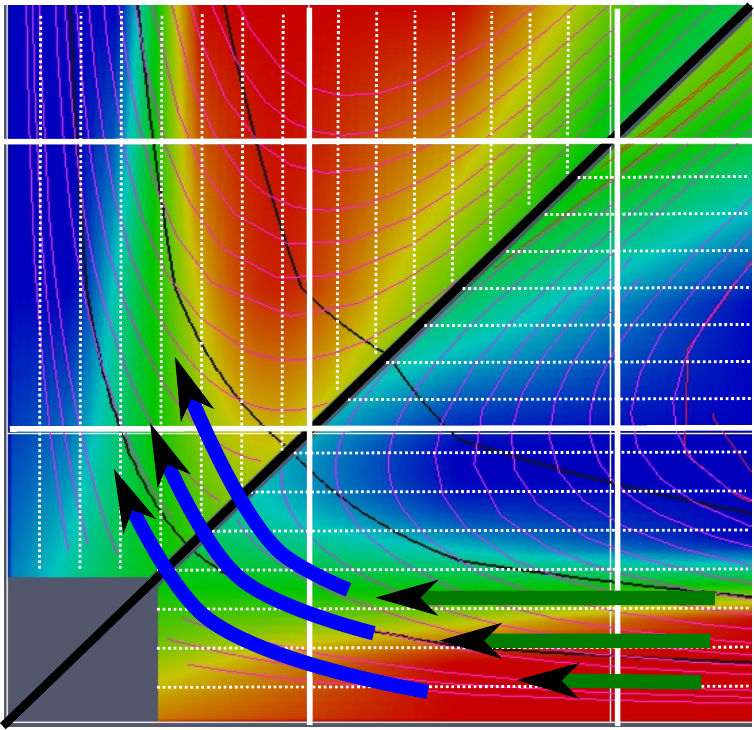} 
                \caption{${Re_B}=3485$, $N_{\rm RSS}=1024$.}
                \label{fig:DuctODTLESN32ReB3500Zoom}
        \end{subfigure}   
        \begin{subfigure}[b]{0.49\textwidth}
		\centering
		\includegraphics[width=0.7\textwidth]
		{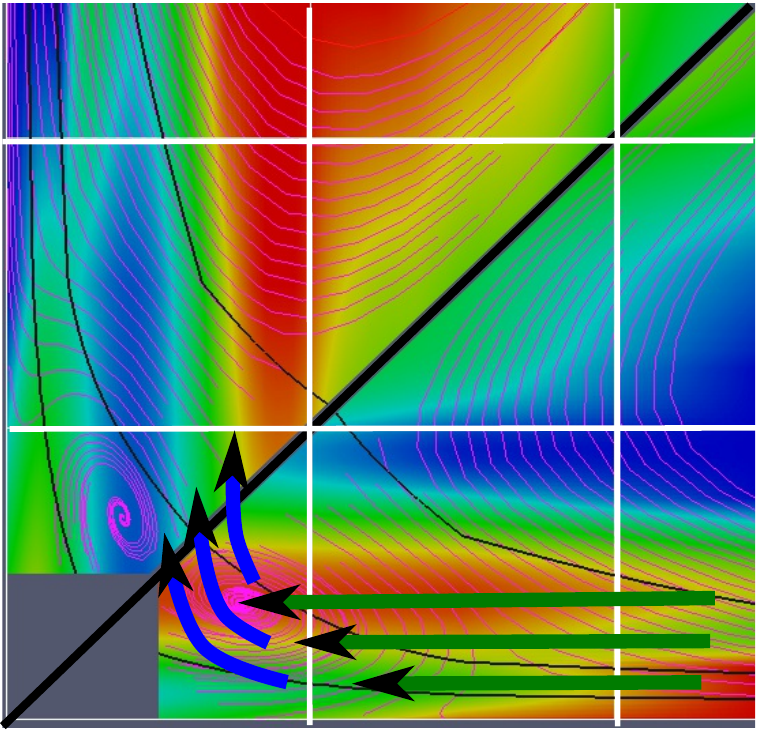}
                \caption{$Re_B=17338$, $N_{\rm RSS}=2048$.}
                \label{fig:DuctODTLESN32ReB17000Zoom}
        \end{subfigure} 	 
       \caption{
Zoom into the corner 
region of figure \ref{fig:DuctODTLESN32ReB3500} ($Re_B=3485$)
and \ref{fig:DuctODTLESN32ReB17000} ($Re_B=17338$). Primary and 
secondary streamlines and vorticity are illustrated similar to figure 
\ref{fig:DuctReB2600}. 
ODTLES properties are illustrated like cell centered (this leads to the gap in 
the flow field at the corner).  \ref{fig:DuctODTLESN32ReB3500Zoom} 
additionally shows the small scale resolution (illustrative with $N_{\rm 
RSS}=256$).
An exemplary cross-stream flow approaches the corner 
parallel to the horizontal wall (green arrows) with wall-normal velocity 
gradients highly resolved only in the XLES-grid $2$ (compare to figure 
\ref{fig:geometryduct2}). 
Near the corner the vertical wall forces a flow stagnation and an associated 
pressure gradient drives the flow in the horizontal direction (blue arrows) 
which is highly resolved only in XLES-grid $3$. For high Reynolds number duct 
flows (\ref{fig:DuctODTLESN32ReB17000Zoom}) both effects occur within one 3D 
cell, which is not well represented by ODTLES because the coupling procedure 
only communicates small scale effects affecting the large scale. This 
could lead to unphysical flow behavior within the 3D cell containing the 
corner.}
                \label{fig:DuctODTLESN32ReB17000Zoom2}
\end{figure}

ODTLES results show a tertiary instability above some threshold Reynolds number 
(between $8446$ and $17338$) in the corner region. 
This model result is not conclusive because the tertiary instability is not 
resolved sufficiently in 3D (see figure \ref{fig:DuctODTLESN32ReB17000Zoom2}). 

The 3D tertiary structure in the corner region is of similar size as the 
corresponding 3D computational cell. Its influence on the flow is primarily 
local.
The insufficient 3D resolution possibly leads to an under-resolved 
pressure gradient which could prevent an adequate change of a fluid parcel's 
direction (blue arrows within figure \ref{fig:DuctODTLESN32ReB17000Zoom2}). 
Nevertheless the cross-stream flow is adequately resolved in the wall-normal 
direction (green arrows within figure \ref{fig:DuctODTLESN32ReB17000Zoom2}). 
More reliable and conclusive results might be possible through future DNS 
studies. 

ODTLES demonstrated its ability to describe and predict non-trivial flow 
behavior including secondary instabilities within a duct flow. 
We suspect that the flow transition leading to tertiary instability is 
a model artifact, although a physical cause cannot be ruled out until 
definitive evidence such as a DNS result becomes available. 
Nevertheless the tertiary structure is very local and not preventing the ODTLES 
model from describing the key flow features of the primary and secondary flow.

\section{ODTLES: Resolution Properties and Efficiency}
\label{s:XLES_ResProps}


Different turbulence models, e.g. RaNS, wall-modeled LES (LES including a 
near-wall 
model), wall-resolved LES (LES with near wall resolution) and ODTLES, differ 
strongly in both represented physical effects and computational effort.
In this section the computational costs of the different model approaches are 
estimated by developing a relation between the grid-size (used as a measure for 
the computational effort) and the Reynolds number following \cite{Chapman1979} 
and especially \cite{Choi:2012} and references cited therein.

The investigated domain is a box of size $L_1 \times L_2 \times L_3$.  A 
highly turbulent boundary layer over a flat-plate airfoil of the 
thickness $\delta$ fills the volume $[x_0, L_1] \times \delta(x_1) \times 
L_3$.
The flow is assumed to reach the plate at $x_1=x_0$ with a turbulent intensity 
$Re_{x_0}$. The boundary layer size $\delta(x_1)$ increases until reaching 
$x_1=L_1$ 
with a corresponding Reynolds number $Re_{L_1}$.
The number of grid cells $N$ within the turbulent 
boundary layer is estimated for ODTLES and compared to RaNS, LES and DNS.



From \cite{Choi:2012} we extract the Reynolds dependent grid size for 
RaNS and wall-modeled LES:
\begin{align}
\label{appeqn:wall-modeledLES}
  N_{RaNS|wm} = 54.7  \frac{L_3}{L_1} n_1 n_2 n_3 Re_{L_1}^{2/7} 
      \left[
	 \left(
	    \frac{Re_{L_1}}{Re_{x_0}}
	 \right)^{5/7} -1
      \right],
\end{align}
for a wall-resolved LES
\begin{align}
\label{appeqn:wall-resolvedLES}
  N_{wr} = 0.021 \frac{L_3}{L_1}  \frac{n_{2,laminar}}{\Delta x_{1,w}^{+} 
\Delta 
x_{3,w}^{+}} 
Re_{L_1}^{13/7} 
      \left[ 1- 
	 \left(
	     \frac{Re_{x_0}}{Re_{L_1}}
	 \right)^{6/7} 
      \right],
\end{align}
and for DNS
\begin{align}
  N_{DNS} = 0.000153 \frac{L_3}{L_1} Re_{L_1}^{37/14}  
        \left[ 1- 
	 \left(
	     \frac{Re_{x_0}}{Re_{L_1}}
	 \right)^{23/14} 
      \right].
\end{align}
Here $n_x n_y n_z$ is the number of grid points within the cube $\delta \times 
\delta \times \delta$, $n_{2,laminar}$ is the number of wall-normal grid points 
within the laminar sublayer, and $x_{k,w}^{+}$ is the LES cell size in wall 
units.

Following \cite{Chapman1979} RaNS typically resolves the cube $\delta \times 
\delta \times \delta$ using $n_x n_y 
n_z \approx 1 \times 20 \times 0.5= 10$ cells, while for wall-modeled LES 
\cite{Choi:2012} report typical grid resolutions $n_x n_y n_z \approx 
[1200,33000]$. 
In wall-resolved LES \cite{Choi:2012} find typical resolution values
$\frac{n_{2,laminar}}{\Delta x_{1,w}^{+} \Delta 
x_{3,w}^{+}} \approx [\frac{1}{390},\frac{1}{25}]$.

Here LES models represent turbulent scales down to the inertial range of 
the 
turbulent cascade. The ODT model, applied within XLES, 
potentially describes the full turbulent cascade within a 1D sub-domain, which 
leaves the 3D grid to capture non-turbulent effects (e.g. the domain or 
secondary instabilities). 
For the flat-plate airfoil even ODT stand-alone potentially leads to reasonable 
results for the case of a turbulent boundary 
layer (\cite{Lignell2012} apply ODT to a comparable turbulent case, but 
including buoyancy, by spatially advancing the ODT line). In consequence the 
(equidistant) XLES 3D resolution $N_k$ in $x_k$-direction is chosen 
independently of the Reynolds number (unless Reynolds-number variations 
triggers 
a global flow structure transition, like the secondary instabilities in a 
turbulent 
duct). 

In the current ODTLES implementation the resolved small scales are represented 
by $N_{\rm RSS_k}$ equidistant cells in $x_k$-direction 
($k=\{1,2,3\}$).
Following \cite{Choi:2012} for highly turbulent flows the number of grid points 
resolving the Kolmogorov length scale along a small distance $\dint x_1$ is 
$N_1= 0.116 
\frac{\dint x_1}{x_1} Re_x^{13/14}$. Equidistant ODTLES uses the smallest 
length scale globally in all 1D sub-domains, leading to
\begin{align}
  N_{ODTLES} = 0.116 K_{ODT} N_1 N_3 \frac{L_3}{L_1} Re_{L_1}^{13/14} 
\end{align}
A factor $K_{ODT} \approx 3 
\times 6$ takes into account that $3$ XLES-grids are used (we assume 
that $N_1=N_2=N_3$) and equidistant ODT uses at least $6$ cells to allow a 
turbulent event (eddy) within the Kolmogorov scale.

In principle ODTLES can be extended to non-equidistant grids within the 1D 
sub-domain, which is for example realized by the adaptive ODT (aODT)
implementation by \cite{Lignell2012}. 
Although adaptive ODT is not used as a sub-grid model within an XLES approach 
yet, 
we investigate this interesting case as a worthwhile perspective and refer 
to it as aODTLES.
For an adaptive grid we assume on average a resolution similar to DNS (in 1D) 
and
integrate over the boundary layer thickness 
with $\frac{\delta}{x}=0.16 Re_x^{-1/7}$ (see \cite{Choi:2012}) in the 1D 
sub-domain, leading to
\begin{align}
N_{aODTLES} = 0.0103936 K_{aODT} N_1 N_3 \frac{L_3}{L_1} Re_{L_1}^{11/14}  
        \left[ 1- 
	 \left(
	     \frac{Re_{x_0}}{Re_{L_1}}
	 \right)^{25/14} 
      \right].
\end{align}
For adaptive ODT, note that we assume $K_{aODT}=3$ because $3$ XLES-grids 
are required (here we assume each XLES-grid uses the same RSS 
resolution and $N_1=N_2=N_3$). 
There is no additional assumption of a minimum number of cells representing the 
Kolmogorov length. 

\begin{figure}
        \centering       
\includegraphics[width=0.6\textwidth]{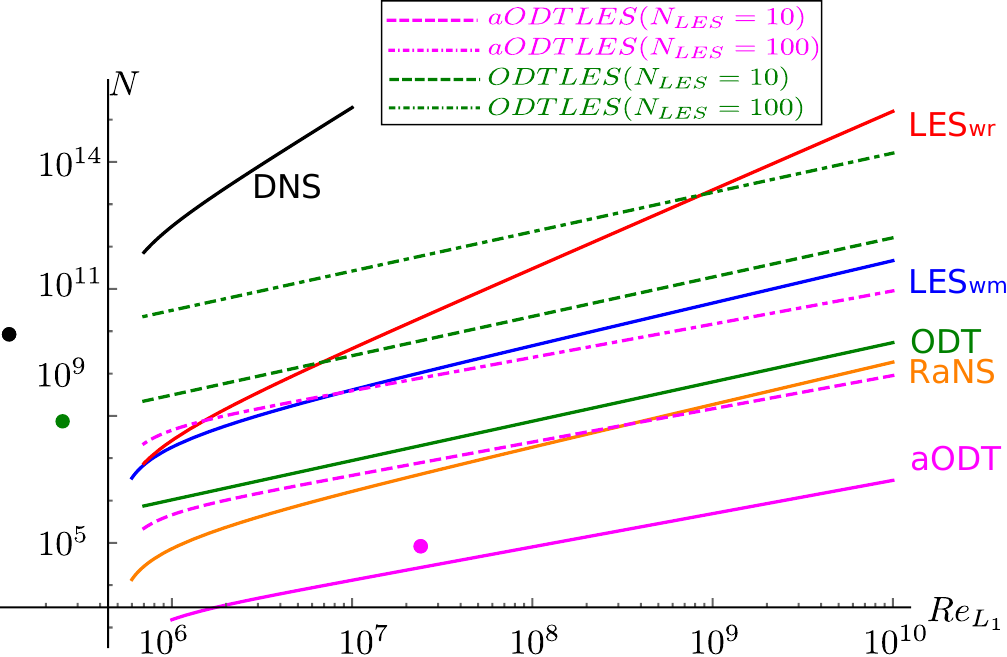}
        \caption{Number of grid points $N$ required for numerical simulations 
of a flow over a flat-plate airfoil with aspect ratio $L_3/L_1=4$ and a 
turbulent 
inflow with $Re_{x_0}=5 \times 10^5$. Wall-modeled LES ($n_1n_2n_3=2500$) 
, wall-resolved LES (${n_{2,laminar}}/{\Delta x_{1,w}^{+} \Delta 
x_{2,w}^{+}}=1/200$), RaNS ($n_1n_2n_3=10$), ODTLES and aODTLES (ODTLES with 
adaptive ODT) 
with $N_1 =N_3 = \{10,100\}$ cells, and the  ODT and adaptive ODT (aODT)  
stand-alone model. Additionally, actual simulation cases for a 
turbulent channel (assuming $Re_{turb} \approx R_{L_1}$) are shown for DNS 
($Re_{\tau}=5200$, black point), ODTLES ($Re_{\tau}=10000$, 
\textcolor{darkgreen}{green point}), and aODT ($Re_{\tau}=6 
\times 10^5$, \textcolor{magenta}{magenta point}). }
\label{figapp:ODTLESefficiency}
\end{figure}

We compare typical RaNS and LES resolutions (following \cite{Choi:2012} and 
\cite{Chapman1979}) with the ODTLES and aODTLES approach for different 3D 
resolutions in figure \ref{figapp:ODTLESefficiency}. Additionally numerical 
computations for a turbulent channel are shown assuming a similarity of the 
turbulent Reynolds number in the channel and 
$Re_{L_1}$. Hereby the DNS by \cite{Moser:2015} (we assume 
$N \approx 8.5\times 10^9$, estimated for $L_3/L_1=4$), an aODT result by 
\cite{Meiselbach2015} 
($N_{aODT} \approx [80000,120000]$ (\cite{MeiselbachEMAIL:2015})) and 
the ODTLES result with $Re_{\tau}=10000$ in section \ref{ss:ODTLESHighRe} are 
used. 

For weakly turbulent flows, ODTLES is subject to additional computational 
costs compared to standard LES.
But ODTLES requires 3D resolution independent of the turbulence intensity 
(except secondary effects), and thus highly turbulent flows in moderately 
complex domains are well described with 
low computational costs. 
In some flow regimes ODTLES is more efficient than wall-modeled LES and 
although it  represents advective and diffusive effects down to the 
Kolmogorov length scale. 
Incorporating adaptive ODT into a XLES 
framework seems to be an especially promising alternative to wall-modeled LES 
and even RaNS 
simulations for highly turbulent flows in simple domains (requiring low 3D 
resolutions).

Note that in the presented estimation the costs of the computation within one 
discrete cell is neglected, because the different modeling strategies typically 
vary by a low factor ($\lesssim 2$), which is not strongly affecting the 
estimation in figure \ref{figapp:ODTLESefficiency}.


\section{Conclusions}
\label{s:Conclusions}

Part I (\cite{Glawe:2014}) introduces XLES, an extended LES approach, 
which is a new strategy to 
simulate complex turbulent flows.
The ODTLES model is one special approach in the XLES family of models, 
employing ODT as a 
sub-grid model. 
XLES in general and especially ODTLES are designed to describe highly turbulent 
flows in domains of moderate complexity. These problems especially occur in 
fundamental research studies of e.g. atmospheric flows and are also relevant in 
engineering.

A previous ODTLES version (by \cite{RC-Schmidt2010} and 
\cite{ED-Gonzalez-Juez2011}) can also be interpreted as ODT closed XLES, but 
with XLES velocities only coupled by a pressure projection. Thus oscillations 
in root mean square velocities occur (reported by \cite{ED-Gonzalez-Juez2011}), 
but without significantly affecting mean profiles. The introduced 
XLES coupling terms directly carry over to scalar properties residing on 
different XLES-grids (e.g. required for a heated duct flow) which is a unsolved 
problem within the previous ODTLES formulation.  

XLES time advances multiple coupled 2D filtered Navier-Stokes realizations, 
each having one Cartesian direction that is highly resolved.
A one-dimensional modeling approach like ODT takes advantage of the 
specific symmetry of a 2D filter.

Especially within ODTLES, the ability of ODT to describe the full turbulent 
spectrum 
allows strongly reduced 3D resolutions 
without corrupting key flow features.

In XLES, separated physical effects (contrary to separated 3D scales in 
LES) are represented by 
appropriate approaches: the 3D resolution represents  
the 3D domain and other physical effects not captured by ODT, e.g. secondary 
instabilities (see section \ref{ss:ODTLESduct}), while turbulent effects, not 
captured by XLES, are 
appropriately represented by the ODT sub-grid model. This includes the 
representation of molecular diffusion and turbulent advection at the 
Kolmogorov length scale (within a 1D sub-domain). 

ODTLES accurately describes a turbulent channel flow up to friction 
Reynolds number $Re \leq 10000$ with high accuracy, 
even with coarse 3D resolution (e.g. $N_{\rm LES} = 16$ cells per direction) 
and is able to reproduce the primary and secondary flow in a square duct with 
similar 3D resolution.

The focus of this work is to introduce the mathematical framework necessary to 
derive the XLES approach, which is one possible way to incorporate ODT into 3D 
simulations. This is implemented here for a simple ODT model to avoid 
e.g. interpolation effects that are introduced by the adaptive ODT model by 
\cite{Lignell2012}. Adaptive ODT outperforms the turbulence 
intensity reachable by equidistant ODT, as \cite{Meiselbach2015} recently 
showed using adaptive ODT simulations up to $Re_{\tau}\leq 6 \times 10^5$.

Incorporating this adaptive ODT model into the XLES filter approach can
potentially serve as an alternative to RaNS simulations in industrial 
applications and additionally include a wide range of small scale physical 
effects, e.g. for flows including Prandtl number effects and 
combustion.

Various additional physical effects, e.g. additional scalar fields, are well 
tested for ODT and can easily be adapted to ODTLES. Furthermore the XLES 
framework is the first approach which consistently couples scalar properties 
and velocities between XLES-grids (in the previous ODTLES version only 
velocities were coupled).
This enables ODTLES to 
compute e.g. fundamental meteorological flows.

Thus ODTLES accurately describes highly turbulent flows including fully 
resolved small scale effects with low computational costs for simple 
domains.

\section*{Acknowledgments}
\label{s:Acknowledgements}

The authors would like to thank H. Kawamura and colleagues and R. Moser and 
colleagues for providing DNS results online (\cite{Kawamura:2013} and 
\cite{Moser:2015}) as well as Philipp Schlatter  and colleagues 
(\cite{Schlatter:2010}) for providing $\tau_W$ statistics for comparison.
This work was 
supported by the Brandenburg University of 
Technology Cottbus-Senftenberg, the Helmholtz graduate research school GeoSim, 
and the Freie Universit\"at Berlin.

\appendix

\section{ODT: Illustrative Results}
\label{app:ODTResults}

As discussed in section \ref{s:XLES2ODTLES}, ODT error terms $\sigma_{\rm ODT}$ 
can roughly be estimated by comparing ODT and DNS flow statistics. 
The turbulent channel is an appropriate study case for ODT because of its 
distinct predominant direction. ODT results with friction Reynolds number 
$Re_{\tau} = 395$ are compared to DNS (\cite{Kawamura:2013}), which allows 
additional comparison to LES-U and  XLES-U results in 
\cite[section 3.4]{Glawe:2014} and ODTLES results in section 
\ref{ss:ODTLESConvergence}.


The ODT model parameters are $C=6.5$ and $Z=300$ which are the same as the 
ODTLES parameters in section \ref{ss:ODTLESConvergence} and 
\ref{ss:ODTLESHighRe}. 
The maximum eddy length $l^{\rm max}$ is chosen to equal the channel 
half height $h$. 
The ODT resolution is $N_{ODT} = 1024$.
To compute reliable ODT flow statistics a larger averaging period (or 
ensemble averaging) is required: the average time is $t_{ave} = 12800$ 
non-dimensional time units after reaching a steady state, which is 
significantly larger than $t_{ave}=20$ in DNS.


\begin{figure}
        \centering
        \begin{subfigure}[b]{0.49\textwidth}
\includegraphics[width=\textwidth]{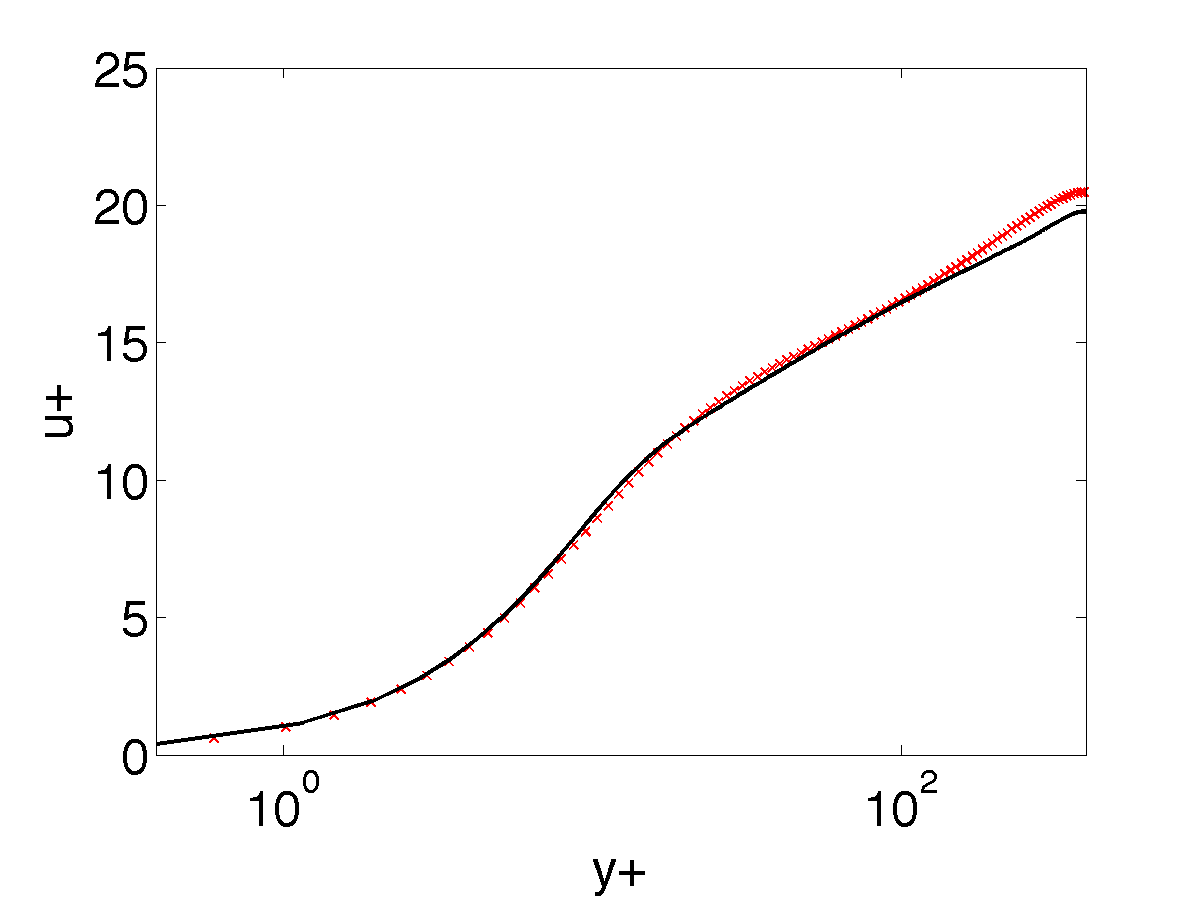}
                \caption{ ODT: \\law of the wall.  }
                \label{fig:umODT}
        \end{subfigure}     
        \begin{subfigure}[b]{0.49\textwidth}
                \includegraphics[width=\textwidth]{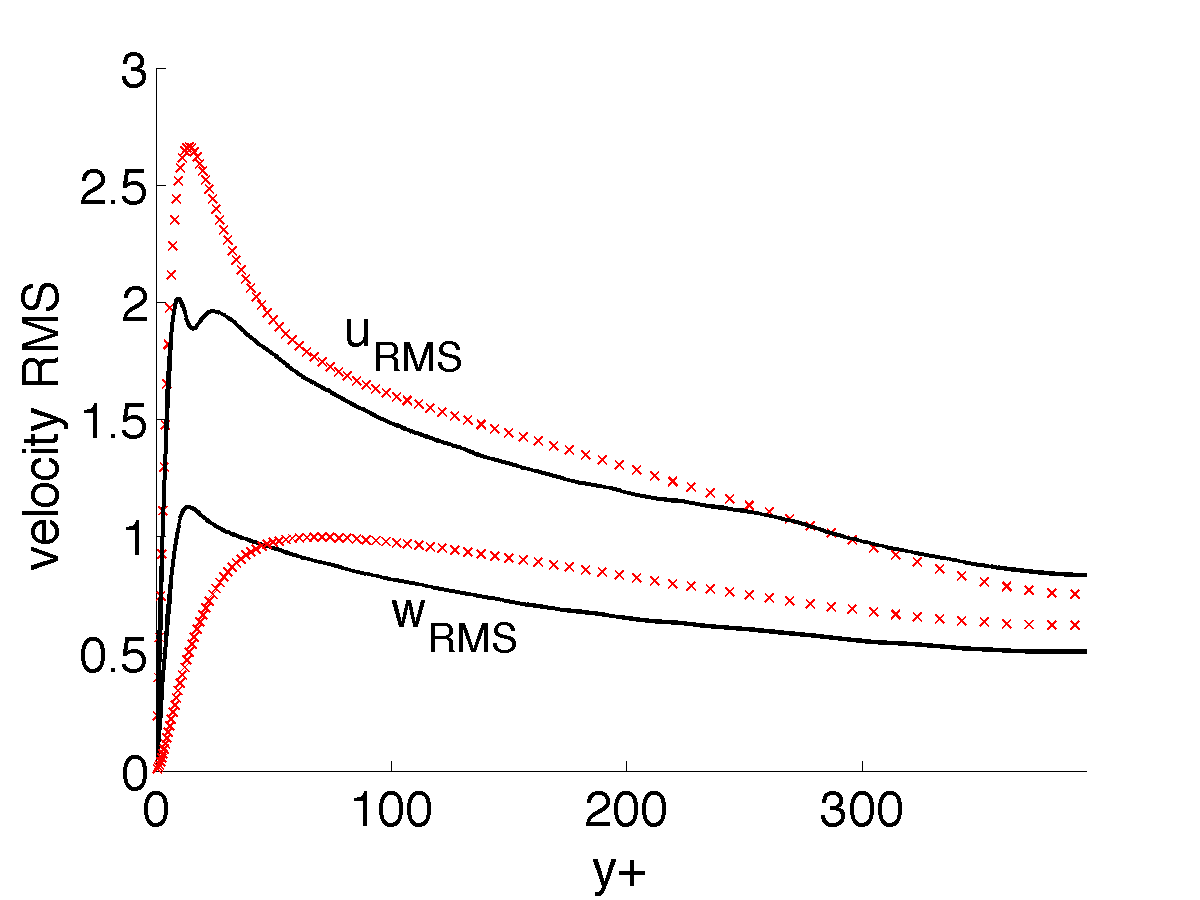}
                \caption{ Streamwise ($u_{\rm RMS}$) and spanwise  
($w_{\rm RMS}$) velocity RMSs.}
                \label{fig:uvwRODT}
        \end{subfigure} 
         \begin{subfigure}[b]{0.49\textwidth}                
\includegraphics[width=\textwidth]
{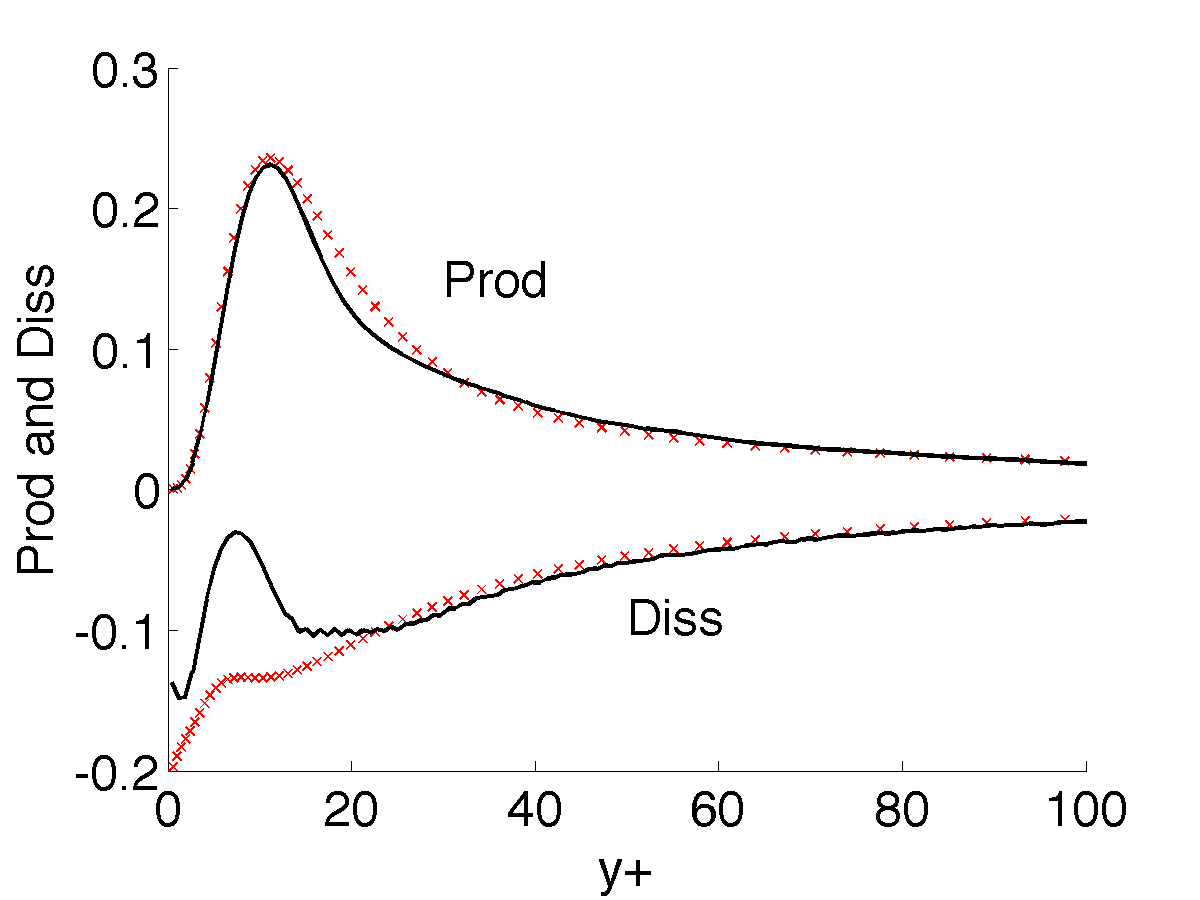}
                \caption{ Production (Prod) and Dissipation (Diss) of the 
turbulent kinetic energy.}
                \label{fig:ProdDissODT}
        \end{subfigure}    
        \begin{subfigure}[b]{0.49\textwidth}
	  \centering               
                \includegraphics[width=0.9\textwidth]
                {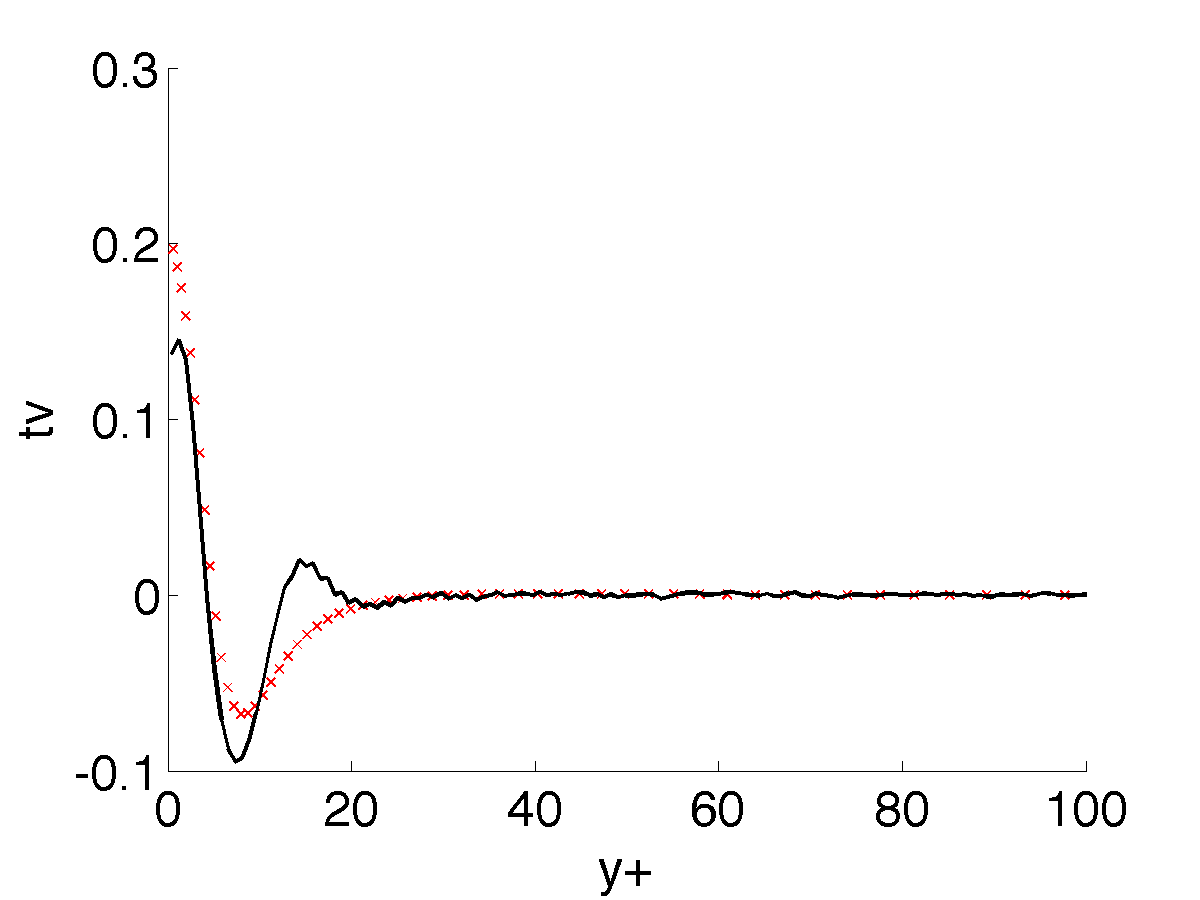}
                \caption{Viscous transport of the turbulent kinetic 
energy (tv).}
                \label{fig:TvODT}
        \end{subfigure}
        \begin{subfigure}[b]{0.49\textwidth}
	  \centering               
                \includegraphics[width=0.9\textwidth]
                {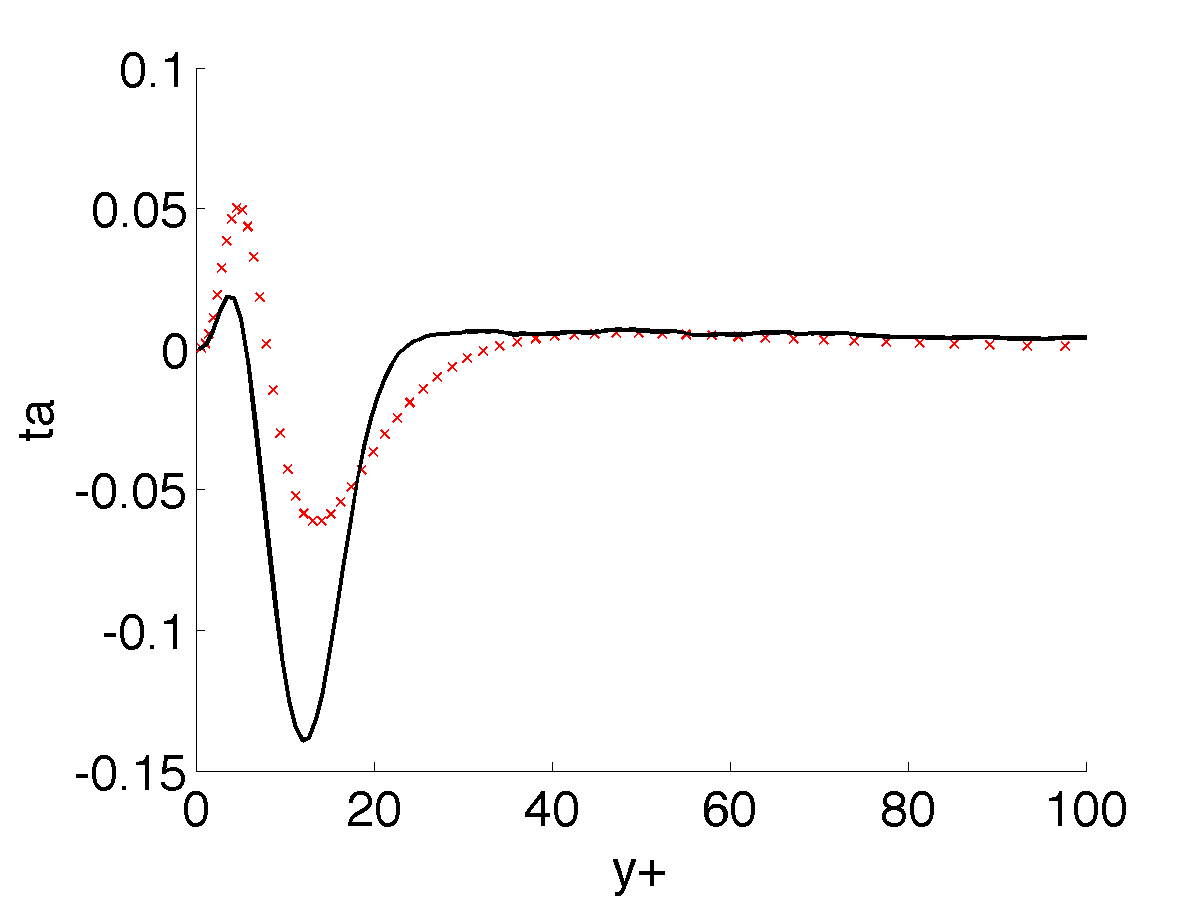}
                \caption{Advective transport of the turbulent kinetic 
energy (ta).}
                \label{fig:TaODT}
        \end{subfigure}         
	    \caption{Turbulent channel flow results ($Re_{\tau}=395$) for DNS 
(\textcolor{red}{small crosses}) and ODT 
(\textcolor{black}{solid}) with $N_{ODT}=1024$ cells.}
	    \label{fig:ODTChannel_Um}
\end{figure}

Figure \ref{fig:ODTChannel_Um} illustrates some representative results for ODT:
The averaged streamwise velocity profile (see figure 
\ref{fig:umODT}) and the overall turbulent kinetic 
energy (see figure \ref{fig:ProdDissODT}--\ref{fig:TaODT}) are described very 
well within the ODT model including the full spectrum of the turbulent 
cascade.
The similarity of ODT and DNS results implies a considerably small ODT model 
error 
$\sigma_{\rm ODT}$.
The ODT computing time is only $\approx 
9$ CPU-seconds to simulate for $t_{ave}=20$ (the overall computing time is 
higher, because 
in the example the flow is averaged over $t_{ave}=12800$).
ODT is a convenient sub-grid 
model because of its low computational costs.
Thereby ODT can compute a wide range of complex physical effects. 
In the turbulent channel case, ODT dynamically produces realistic wall 
profiles (see section \ref{ss:ODTeffects}).
These properties outperform commonly used eddy viscosity models. 

Further ODT results, including various physical small scale effects, are 
available in the literature (e.g. see 
\cite{AR-Kerstein2001}, \cite{HeikoSchmidt:2013}, \cite{Wunsch:2005}, 
and \cite{Schulz:2013}).

\section{ODTLES: Numerical Implementation}
\label{app:NumImpl}

Table \ref{tab:NumericalSchemes} summarizes the numerical discretizations of 
the single fractional steps within the ODTLES advancement cycle (section 
\ref{ss:TimeAdvancement}). The ODT advancement (Eq. 
(\ref{eqn:timeIntegral5})) contains the XLES diffusion terms 
($\mathcal{\vec{D}}_{\rm XLES}$) treated as forcing terms for ODTLES.

\begin{table}[ht]
\centering
\caption{Numerical Schemes:\\
 In time:
 EE1 (1st order explicit Euler), RK3 (3rd order Runge-Kutta), CN (2nd order 
Crank-Nicolson), IE1 (1st order implicit Euler).\\
In space:
UP1 (1st order upwind), CDM (2nd order central difference method).
}
\label{tab:NumericalSchemes}
  \begin{tabular}{|l || l | l |}
  \hline
 factional step   	 & time scheme & spatial scheme   \\ 
\hline \hline
  Eq. (\ref{eqn:timeIntegral})   & EE1 &  UP1
  \\ \hline
  Eq. (\ref{eqn:timeIntegral2})   & RK3-CN + RK3-RK3 &  CDM
  \\ \hline
  Eq. (\ref{eqn:timeIntegral3})   & EE1 &  CDM
  \\ \hline
    Eq. (\ref{eqn:timeIntegral5})   & IE1 + triplet map + EE1&  CDM
  \\ \hline
    Eq. (\ref{eqn:timeIntegral6})   & EE1 &  CDM
  \\ \hline
\end{tabular}
\end{table}

The coupled advection schemes RK3-CN-CDM and RK3-RK3-CDM are described and 
validated in part I \cite[section 3.3]{Glawe:2014}.
All velocities are discretized using a staggered grid.



%

 \bibliographystyle{elsarticle-harv} 
 \bibliography{glawe.bib}


%
%
%
\end{document}